\documentclass[fleqn,usenatbib]{mnras}

\usepackage{newtxtext,newtxmath}

\usepackage[T1]{fontenc}
\usepackage{ae,aecompl}

\usepackage{graphicx}	
\usepackage{amsmath}	
\usepackage{subcaption}
\usepackage{cleveref}
\usepackage{multirow}
\usepackage{placeins}

\DeclareMathOperator{\Tr}{Tr}

\title[Time Dependent Global 21cm Modelling]{Use of Time Dependent Data in Bayesian Global 21cm Foreground and Signal Modelling}

\author[D. Anstey et al.]{
Dominic Anstey,$^{1,2}$\thanks{E-mail: \href{da401@mrao.cam.ac.uk}{da401@mrao.cam.ac.uk}}
Eloy de Lera Acedo$^{1,2}$\thanks{E-mail: \href{eloy@mrao.cam.ac.uk}{eloy@mrao.cam.ac.uk}}
 and Will Handley$^{1,2}$\thanks{E-mail: \href{wh260@mrao.cam.ac.uk
}{wh260@mrao.cam.ac.uk}}
\\
$^{1}$Astrophysics Group, Cavendish Laboratory, J. J. Thomson Avenue, Cambridge, CB3 0HE, UK\\
$^{2}$Kavli Institute for Cosmology, Madingley Road, Cambridge, CB3 0HA, UK\\
}

\date{Accepted XXX. Received YYY; in original form ZZZ}

\pubyear{2022}

\begin{document}
\label{firstpage}
\pagerange{\pageref{firstpage}--\pageref{lastpage}}
\maketitle

\begin{abstract}
Global 21cm cosmology aims to investigate the cosmic dawn and epoch of reionisation by measuring the sky averaged HI absorption signal, which requires, accurate modelling of, or correction for, the bright radio foregrounds and distortions arising from chromaticity of the antenna beam. We investigate the effect of improving foreground modelling by fitting data sets from many observation times simultaneously in a single Bayesian analysis, fitting for the same parameter set by performing these fits on simulated data. We find that for a hexagonal dipole antenna, this simultaneous fitting produces a significant improvement in the accuracy of the recovered 21cm signal, relative to fitting a time average of the data. Furthermore, the recovered models of the foreground are also seen to become more accurate by up to a factor of $\sim$2-3 relative to time averaged fitting. For a less chromatic log spiral antenna, no significant improvement in signal recovery was found by this process. However, the modelling of the foregrounds was still significantly improved. We also investigate extending this technique to fit multiple data sets from different antennae simultaneously for the same parameters. This is also found to improve both 21cm signal and foreground modelling, to a higher degree than fitting data set from multiple times from the same antenna. 
\end{abstract}




\section{Introduction}\label{sec:intro}
The period of cosmic history between redshifts $\sim$ 5 and 50, covering the Dark Ages, the Cosmic Dawn when the first stars formed, and the Epoch of Reionisation, is currently not well understood. Therefore, the investigation of this period is an important goal of cosmology. One of the most promising mechanisms of probing this period of cosmic history is provided by the absorption from and emission into the background radiation field by neutral hydrogen. 

HI has a hyperfine transition at 21cm. This means it can absorb from or emit into the background radiation field at that wavelength. However, several astrophysical properties can alter the populations of HI atoms in the upper and lower energy levels of this transition, and so affect the degree of absorption. The most critical of these is the Wouthuysen-Field effect \citep{wouthuysen52, field58}, by which Lyman-$\alpha$ radiation couples the spin temperature of HI gas to its kinetic temperature. As a result, changes in background temperature with frequency, with appropriate redshifting from 21cm, can be used to investigate changes in astrophysical properties such as Lyman-$\alpha$, radiation and gas temperature, as well as other properties such as gas collisions and ionisation, that also affect the absorption \citep{furlanetto16}. 

The all-sky average of this absorption signal, can, in theory, be detected by a single wide-beam antenna. Several such `Global' 21cm experiments are currently working on detecting this signal, such as REACH, EDGES \citep{bowman08}, SARAS \citep{patra13,singh18}, DAPPER (\href{https://www.colorado.edu/ness/dark-ages-polarimeter-pathfinder-dapper}{https://www.colorado.edu/ness/dark-ages-polarimeter-pathfinder-dapper}), SCI-HI \citep{voytek14}, LEDA \citep{price18}, PRIZM \cite{philip19}, and MIST (\href{http://www.physics.mcgill.ca/mist/}{http://www.physics.mcgill.ca/mist/}). 

However, one of the primary challenges with detecting this global 21cm signal is the presence of bright radio foregrounds, predominantly galactic synchrotron and free-free radiation, that can exceed the expected brightness of the 21cm signal by 3-4 orders of magnitude \citep{shaver99}. These foregrounds are expected to be spectrally much smoother than the signal, owing to their predominantly power law nature, and so could be distinguished from the signal by fitting them with smooth functions such as polynomials \citep{sathyanarayana15, bevins21, bowman18}. However, this is made more difficult by chromatic distortion. Measuring the global 21cm signal requires a wide-band antenna, for which some degree of variation of the antenna gain pattern with frequency is very difficult to avoid. Such chromatic variations will couple to the bright radio foregrounds and introduce chromatic distortions into the resulting data, which are non-smooth. These chromatic distortions can easily mask or distort the detected signal if not accurately removed or modelled \citep{tauscher20a, anstey21}. 

Therefore, accurately determining the radio foregrounds and associated chromatic distortions is highly important in achieving a detection of the global 21cm signal. Two of the key pipelines currently under development to achieve this are a Bayesian nested sampling algorithm based on parameterised, physical-motivated modelling of the foregrounds, presented in \citet{anstey21}, and an algorithm based on Singular Value Decomposition of foreground and chromaticity training sets, presented in \citet{tauscher18}.

The follow-up paper \citet{tauscher20b}, to the SVD pipeline presented in \citet{tauscher18} included an analysis of the effects of time variance of radio foregrounds and chromatic distortions due to the Earth's rotation on that pipeline and found that exploiting this time variance enabled improved modelling of the foregrounds and so improved signal recovery.

In this paper, we analyse the effects of time variation of the foreground on the physically parameterised nested sampling pipeline presented in \citet{anstey21} and demonstrate how this pipeline can exploit this variance to improve foreground modelling and signal recovery, with applicability to global 21cm experiments in general.

In \Cref{sec:methods}, we discuss the method by which time varying foregrounds and chromaticity can be properly incorporated into the pipeline. In \Cref{comparison} we compare the results of fitting time separated data bins to fits of time averaged data, to quantify the effects exploiting time variation can have. In \Cref{variations}, we investigate the impact that the number of time bins used and the LST range they cover has on the ability to recover the 21cm signal. In \Cref{sec:multi_ant}, we present an extension to this method in which data sets from multiple different antennae can be used simultaneously, taking advantage of the change in foregrounds and chromaticity between antennae as well as between time bins. In \Cref{sec:conclusions}, we present our conclusions.

\section{Methods}\label{sec:methods}
In a Global 21cm experiment, the sky-averaged monopole 21cm signal is expected to be approximately uniform across the entire sky. The foregrounds, however, are spatially dependent, and will therefore change with the rotation of the Earth. 

One of the ways in which foregrounds can be modelled is as a general parameterised function, such as the polynomials foregrounds used in \citet{bowman18}, \citet{singh18} and \citet{bevins21}. In models such as this, the optimum parameter values can be expected to change as the foregrounds change. This means that if observation data is split into separate time bins, each bin will require a distinct foreground model.  

\begin{figure}
    \centering
    \includegraphics[width=\columnwidth]{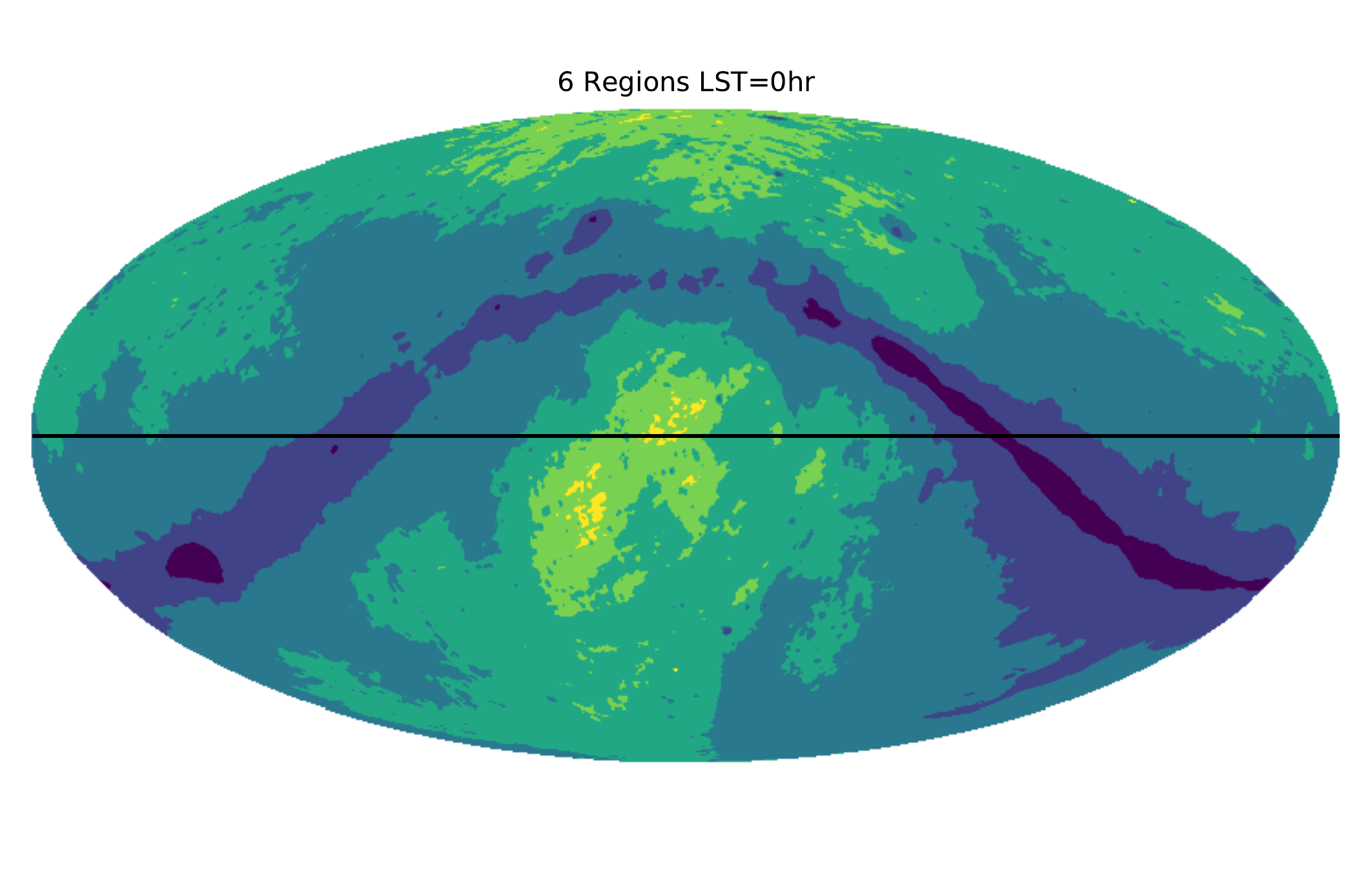}
    \hfill
    \includegraphics[width=\columnwidth]{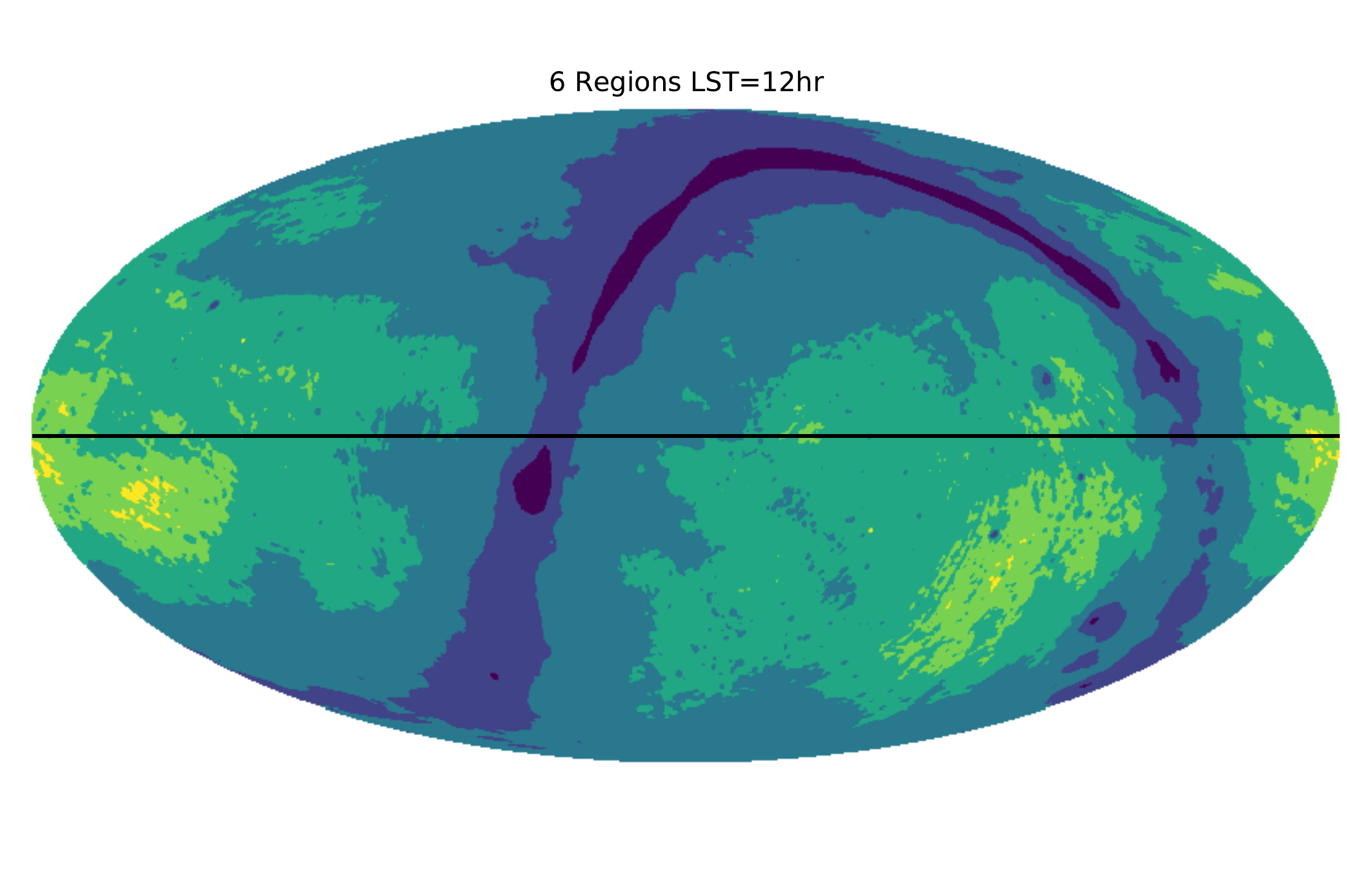}
    \caption{Plots of the division of the sky into 6 regions of similar spectral index for local sidereal times 0hrs and 12hrs, for an antenna located in the Karoo radio reserve. The black line shows the horizon.}
    \label{fig:region_LST}
\end{figure}

However, the foreground could also be modelled based on a spatially-dependent physical property of the sky. If such a parameterisation scheme is used, then the values of the foreground parameters are independent of the observing time despite the foregrounds in general being time dependent, as demonstrated in \Cref{fig:region_LST}, and so a model fit to data at any observation time should converge on the same parameter values. The result of this is that it is possible to take multiple data sets at different observing times, and fit them jointly to corresponding models using the same parameter samples for all. In this work, we investigate how this effect might be exploited in order to achieve more accurate foreground modelling and thus a more accurate reconstruction of the 21cm signal. 

For this work, we use the parameterisation based on foreground spectral index presented in \citet{anstey21}, hereafter A21. In this model, the sky is subdivided into a number $N$ regions, within which the spectral index is modelled as constant. The unknown values of the spectral indices serve as the parameters, with a simulated foreground data generated using this parameterised spectral index map giving the parameterised foreground model.

The A21 pipeline is based on the Bayesian nested sampling algorithm \texttt{PolyChord} \citep{handley15a,handley15b}. In order to achieve such a joint fit in practice in a Bayesian data analysis pipeline, the multiple data sets need to be incorporated into the same likelihood. In A21, a Gaussian likelihood of the form
\begin{multline}\label{eq:integ_L}
    \log\mathcal{L} = \sum_{i}-\frac{1}{2}\log\left(2\pi\sigma_\mathrm{n}^{2}\right) \\ - \frac{1}{2}\left(\frac{T_\mathrm{data}\left(\nu_{i}\right)-\left(T_\mathrm{F}\left(\nu_{i}, \theta_\mathrm{F}\right) + T_{S}\left(\nu_{i},\theta_\mathrm{S}\right)\right)}{\sigma_\mathrm{n}}\right)^{2}
\end{multline}
was used, where $T_\mathrm{data}\left(\nu_{i}\right)$ refers to the observation data and $T_\mathrm{F}\left(\nu_{i}, \theta_\mathrm{F}\right)$ and $T_\mathrm{S}\left(\nu_{i}, \theta_\mathrm{S}\right)$ refer to the foreground and signal models respectively, with parameters $\theta_\mathrm{F}$ and $\theta_\mathrm{S}$. $\sigma_\mathrm{n}$ is an additional parameter for the Gaussian noise. In this case, the data and foreground model are both integrated over time, which can be written more explicitly as 
\begin{multline}\label{eq:integ_L_exp}
    \log\mathcal{L} = \sum_{i}-\frac{1}{2}\log\left(2\pi\sigma_\mathrm{n}^{2}\right) \\ - \frac{1}{2}\left(\frac{\frac{1}{N_\mathrm{t}}\sum_j\left[T_\mathrm{data}\left(\nu_{i}, t_j\right)\right]-\left(\frac{1}{N_\mathrm{t}}\sum_j\left[T_\mathrm{F}\left(\nu_{i}, t_j, \theta_\mathrm{F}\right)\right] + T_{S}\left(\nu_{i},\theta_\mathrm{S}\right)\right)}{\sigma_\mathrm{n}}\right)^{2}
\end{multline}
for time bins $t_j$. However, if the foreground model $T_\mathrm{F}\left(\nu_{i}, \theta_\mathrm{F}\right)$ is such that the posterior values of $\theta_\mathrm{F}$ are expected to be the same for any individual time bin, as discussed above, the likelihood can instead be formulated as 
\begin{multline}\label{eq:sep_L}
    \log\mathcal{L} = \sum_{i}\sum_{j}-\frac{1}{2}\log\left(2\pi\sigma_\mathrm{n}^{2}\right) \\ - \frac{1}{2}\left(\frac{T_\mathrm{data}\left(\nu_{i}, t_{j}\right)-\left(T_\mathrm{F}\left(\nu_{i}, t_{j}, \theta_\mathrm{F}\right) + T_\mathrm{S}\left(\nu_{i},\theta_\mathrm{S}\right)\right)}{\sigma_\mathrm{n}}\right)^{2},
\end{multline}
which fits each data time bin simultaneously to its corresponding model to inform a single shared set of parameters, rather than fitting an integrated data set to a single model.

In this work, we will analyse the effect of using this likelihood, henceforth referred to as 'time-separated', relative to the 'time-integrated' likelihood in \Cref{eq:integ_L_exp}, as well as the effects of changing the number and range of time bins used.

\subsection{Likelihood Calculation}\label{sec:calc_speed}
The time required to evaluate an instance of the likelihood in \Cref{eq:sep_L} for a given parameter sample set grows linearly with the number of time bins $N_\mathrm{t}$. Therefore, fits using this likelihood can be performed significantly faster if the summation over time bins can be precalculated outside of the likelihood. 

For the simple Gaussian likelihood used here, this can be implemented given two conditions:
\begin{itemize}
    \item The foreground model can be expressed as a product of a parameter-independent component and a parameter-dependent component, where the parameter-dependent component is independent of the observing time:
    \begin{equation}
        T_\mathrm{F}\left(\nu_{i}, t_{j}, \theta_\mathrm{F}\right) = K\left(\nu_i, t_j\right)\times F\left(\nu_i, \theta_\mathrm{F}\right).
    \end{equation}
    where $K\left(\nu_i, t_j\right)$ is parameter ($\theta_\mathrm{F}$) independent and $F\left(\nu_i, \theta_\mathrm{F}\right)$ is parameter dependent. 
    
    \item The noise has no parameterised time-dependence. This requires either that the noise parameter $\sigma_\mathrm{n}$ is independent of time, or that it can be expressed as a product of a parameter-independent component and a parameter-dependent component, where the parameter-dependent component is independent of the observing time, as for the foregrounds:
    \begin{equation}
        \sigma_\mathrm{n}\left(\nu_{i}, t_{j}, \theta_\sigma\right) = f\left(\theta_\sigma\right) \times g\left(\nu_i, t_j\right).
    \end{equation}
\end{itemize}
The first condition is satisfied by the foreground model from A21 by definition, which uses a foreground model of the form
\begin{equation}
 T_\mathrm{F}\left(\nu_{i}, t_{j}, \theta_\mathrm{F}\right) = \sum_k K_k\left(\nu_i, t_j\right)F\left(\nu_i, \theta_{\mathrm{F}\,k}\right).   
\end{equation}

The second condition, however, may not be true in practice. Therefore, this method of improving the fitting speed may not be possible on real data. However, in this paper, all simulated data sets are given the same magnitude of noise for each data bin, such that $\sigma_\mathrm{n}$ is independent of time. This satisfies the second condition by design and thus allowing this precalculation to be used. The method by which this is implemented is to express the likelihood as 
\begin{multline}
        \log\mathcal{L} = -\frac{1}{2} N_\mathrm{t} N_{\nu} \log\left(2\pi \sigma_\mathrm{n} \right) \\ -\frac{D}{2{\sigma_\mathrm{n}}^2}  - \frac{1}{{2\sigma_\mathrm{n}}^2} \left[\sum_i \sum_k K_{\mathrm{sq}\, ik} {F\left(\nu_i, \theta_{\mathrm{F}\, k}\right)}^2 + \right.\\ \left. \sum_{k_1} \sum_{k_2}\left[ \sum_i K_{\mathrm{cross}\, i\,k_1\, k_2} F\left(\nu_i, \theta_{\mathrm{F}\, k_1}\right) F\left(\nu_i, \theta_{\mathrm{F}\, k_2}\right) \right] \right.\\ \left. - \Tr{\sum_i K_{\mathrm{cross}\, i\,k_1\, k_2} F\left(\nu_i, \theta_{\mathrm{F}\, k_1}\right) F\left(\nu_i, \theta_{\mathrm{F}\, k_2}\right)} \right] \\
        - \frac{1}{{\sigma_\mathrm{n}}^2} \sum_i \left[\frac{N_\mathrm{t}}{2} {T_\mathrm{S}\left(\nu_{i},\theta_\mathrm{S}\right)}^2
        - \sum_k T_{{\mathrm{D} \cdot \mathrm{K}}\, ik} F\left(\nu_i, \theta_{\mathrm{F}\, k}\right) \right.\\ \left. - T_{\mathrm{D}\, i}T_\mathrm{S}\left(\nu_{i},\theta_\mathrm{S}\right) + \sum_k K_{k}F\left(\nu_i, \theta_{\mathrm{F}\, k}\right)T_\mathrm{S}\left(\nu_{i},\theta_\mathrm{S}\right)\right],
\end{multline}
where $N_{\nu}$ and $N_\mathrm{t}$ are the number of frequency and time bins respectively and
\begin{itemize}
    \item $D = \sum_i \sum_j {T_\mathrm{data}\left(\nu_{i}, t_{j}\right)}^2$
    \item $T_{\mathrm{D}\, i} = \sum_j T_\mathrm{data}\left(\nu_{i}, t_{j}\right)$
    \item $K_{ik} = \sum_j K_k\left(\nu_i, t_j\right)$
    \item $T_{{\mathrm{D} \cdot \mathrm{K}}\, ik} = \sum_j T_\mathrm{data}\left(\nu_{i}, t_{j}\right) K_k\left(\nu_i, t_j\right)$
    \item $K_{\mathrm{sq}\, i\,k} = \sum_j {K_k\left(\nu_i, t_j\right)}^2$
    \item $K_{\mathrm{cross}\, i\,k_1\, k_2} = \sum_j K_{k_1}\left(\nu_i, t_j\right) K_{k_2}\left(\nu_i, t_j\right)$
\end{itemize}
which can be precalculated outside the likelihood. This removes all time integration from within the likelihood evaluation.

\subsection{Optimum Foreground Parameters}\label{sec:reg_div}
As described in A21, the foreground model used in this work has a variable complexity based on the number of regions $N$ that the sky is subdivided in to. Increasing the number of regions allows the foreground model to be more detailed and potentially fit for the true foregrounds more accurately, but at the cost of requiring additional parameters that give additional freedom to the model, which could risk fitting away the 21cm signal as part of the foreground.

It is therefore important to find the optimum number of foreground parameters to use. We find this optimum by calculating the Bayesian evidence of model fits. The Bayesian evidence is proportional to the probability of the model given the data, marginalised over all possible parameter values of the model. This naturally implements an Occam's Razor effect, in which models that have more parameters but do not give any improvement to the fit have a lower evidence. Therefore, the optimal number of parameters can be found by fitting models with a range of $N$s. The one with the highest evidence will then be the model that gives the best fit to the data with the fewest parameters. 

\section{Comparison of Processes}\label{comparison}
\subsection{Hexagonal Dipole}\label{comparison_hex_dipole}

We first test the effects of using time-separated fitting in comparison to time-averaged fitting. In order to perform this comparison, we first generated some sets of simulated global 21cm experiment observation data according to
\begin{multline}\label{eq:sim_data_gen}
        T_\mathrm{data}\left(\nu, t\right) = \frac{1}{4\pi}\int_{0}^{4\pi}D\left(\Omega,\nu\right)\times \\ \left[\left(T_\mathrm{230}\left(\Omega, t\right) - T_\mathrm{CMB}\right)\left(\frac{\nu}{230}\right)^{-\beta\left(\Omega \right)} + T_\mathrm{CMB}\right]d\Omega + \hat{\sigma},
\end{multline}
where $T_\mathrm{CMB}$ is the cosmic microwave background temperature, set as 2.725K, $\beta\left(\Omega\right)$ is the spectral index map described in A21, $T_\mathrm{230}\left(\Omega, t\right)$ is an instance of the 2008 Global Sky Model \citep{deoliveiracosta08} at 230MHz rotated appropriately for an antenna located in the Karoo radio reserve at the observing time $t$, and $\hat{\sigma}$ is Gaussian noise, which we set as having a $\sigma$ of 0.1K for each time bin. This noise level was chosen to be large enough to interfere with signal detection for a single time bin, but small enough to reduce to a level much lower than the signal for the numbers of time bins used in the following test cases. $D\left(\Omega,\nu\right)$ is the directivity pattern of an antenna, which we take as a hexagonal-bladed dipole antenna, as is used in the experiment REACH \citep{mission}. 

\begin{figure}
    \centering
    \includegraphics[width=\columnwidth]{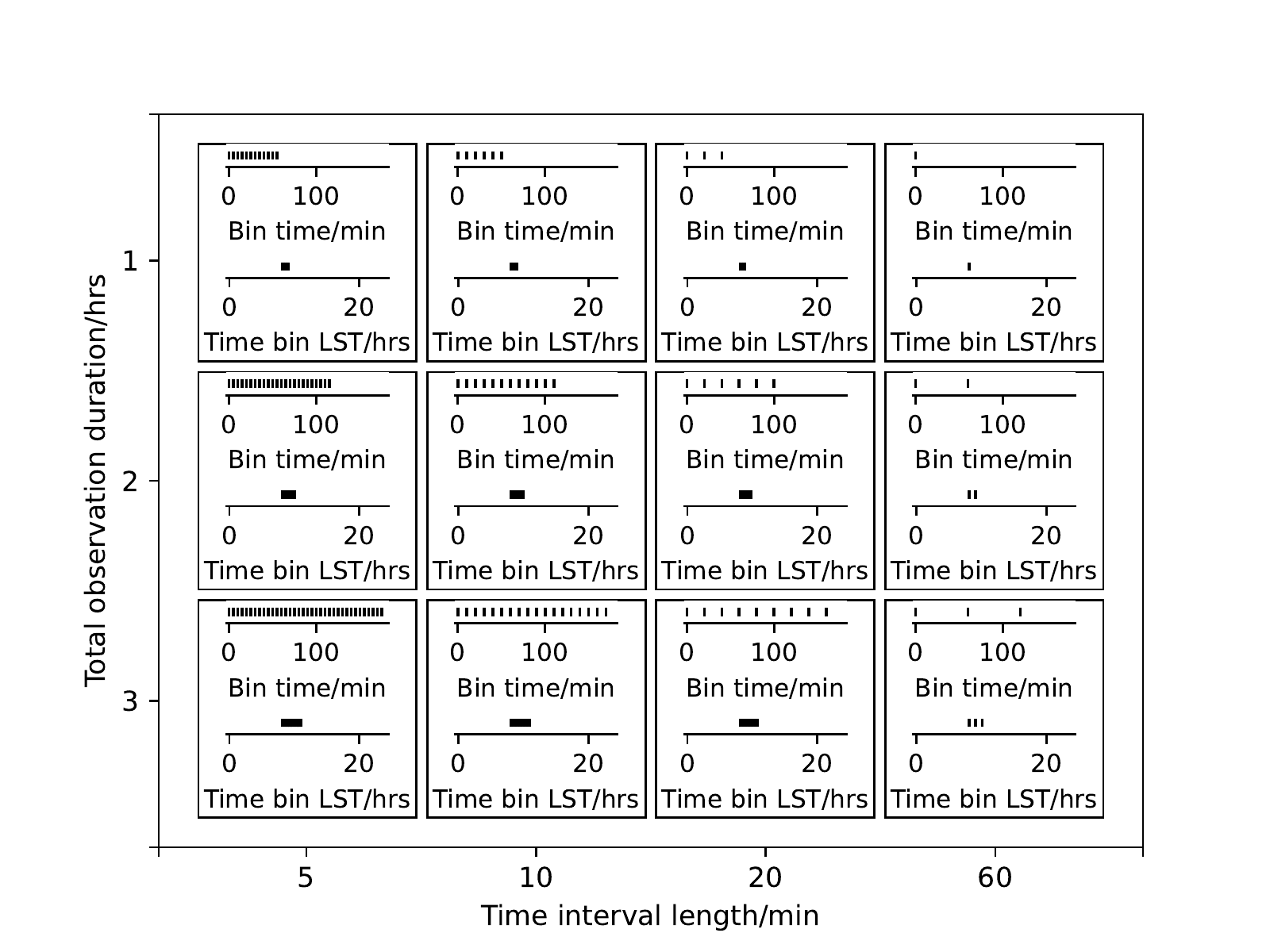}
    \caption{Time bin layouts for tests comparing time-averaged and time-separated fitting, showing the total observation time of the setup on the y-axis and the interval between time bins on the x-axis. In each subplot, the upper plot shows the time of each time bin after the start time of 00:00:00 01-01-2019 UTC and the lower plot shows the LSTs of those time bins.}
    \label{fig:comp_bins}
\end{figure}

We generated 12 simulated data sets, for 1, 2 and 3 hours of observation time, with 5, 10, 20 and 60 minutes between successive bins in each case. In all cases, we took the observation to begin at 00:00:00 01-01-2019 Universal Coordinated Time (UTC), which is at a Local Sidereal Time (LST) of 8.12 hours for an antenna in the Karoo radio reserve. The full layout of time bins for each of the 12 cases are shown in \Cref{fig:comp_bins}

For each of these data sets, we injected a mock Gaussian 21cm signal with a centre frequency 80MHz, a width of 15MHz and an amplitude of 0.155K to every time bin. We then attempted to jointly fit for the foregrounds and signal using the A21 modelling process, fitting both the separated time bins using the likelihood given in \Cref{eq:sep_L} and an average of the time bins of the set using the likelihood given in \Cref{eq:integ_L}. These fits were performed for a range of $N$s in order to identify the optimum number of parameters that gave the highest evidence in each case. The signal model was taken as a Gaussian with three parameters, the centre frequency $f_0$, the width $\sigma$, and the amplitude $A$.

\begin{figure}
    \centering
    \includegraphics[width=\columnwidth]{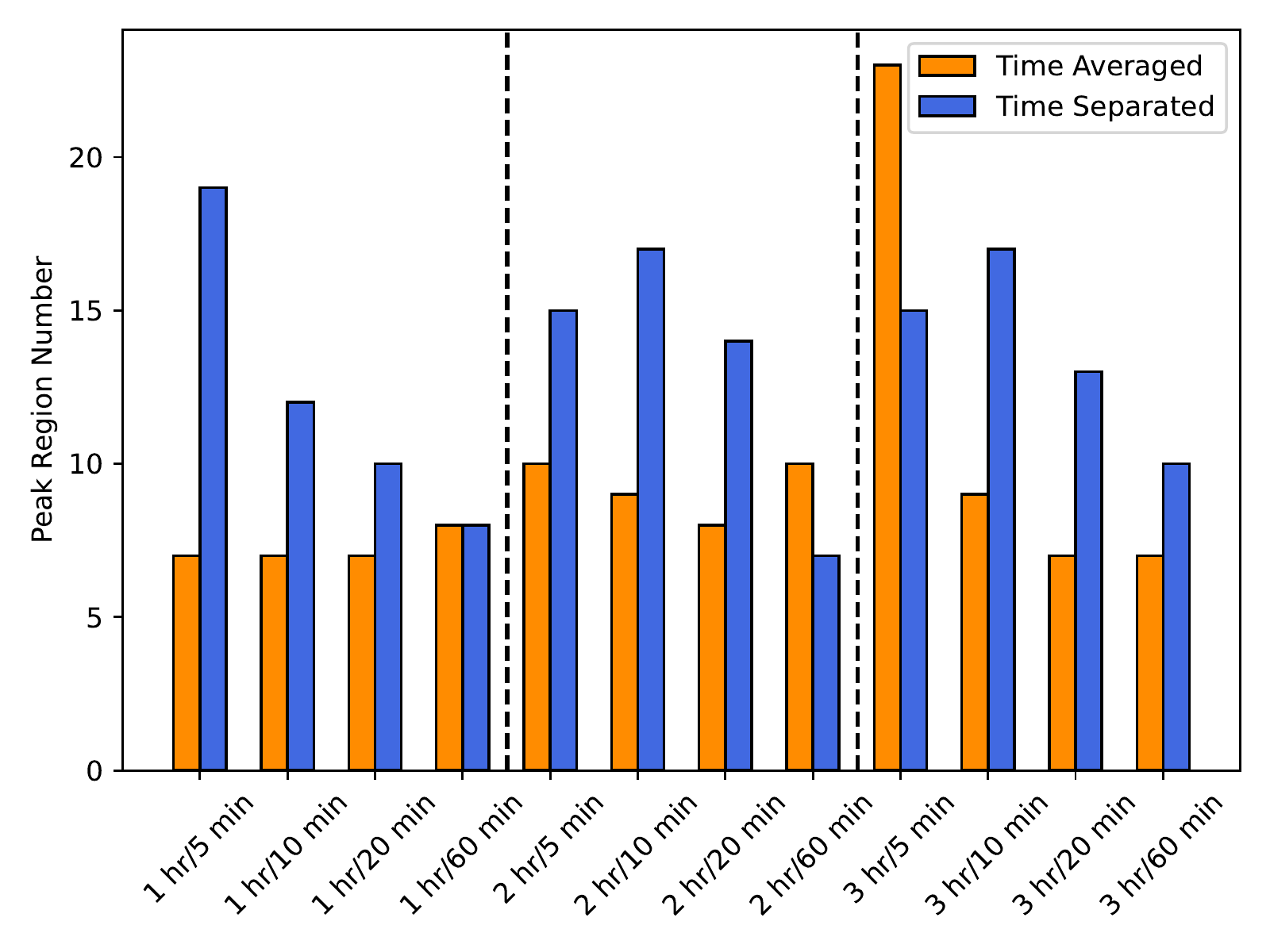}
    \caption{Plot of the number of regions $N$ that the foreground model was divided into in order to give the highest evidence model fit for each of the simulated data sets shown in \Cref{fig:comp_bins}, generated using a hexagonal dipole antenna.}
    \label{fig:comp_peak_regions_hex}
\end{figure}

The number of regions required by the foreground model to give the highest evidence in each case are compiled in \Cref{fig:comp_peak_regions_hex}. It can be seen from these results that for the majority of cases, the time separated fit peaks at a significantly higher number of regions than the time averaged fit. This effect, and the consequences it has on foreground reconstruction, will be discussed in more detail later in this work.

In order to quantify the degree of difference in the quality of the recovered signals for the two methods, we use the Kullback–Leibler divergence (DKL). The DKL quantifies the relative information of different probability distributions \citep{kullback51, handley19}. Therefore, by calculating the DKL of the posterior distribution of the three signal parameters of each fitted model, marginalising over the foreground and noise parameters, the relative information contained in the recovered signals can be quantified. This provides a measure of how well each method of fitting has recovered the signal. In order to calculate this, we use the kernel density estimator in the analysis software MARGARINE \citep{bevins22a, bevins22b}.

\begin{figure*}
\centering
\begin{subfigure}[b]{\columnwidth}
\includegraphics[width=\columnwidth]{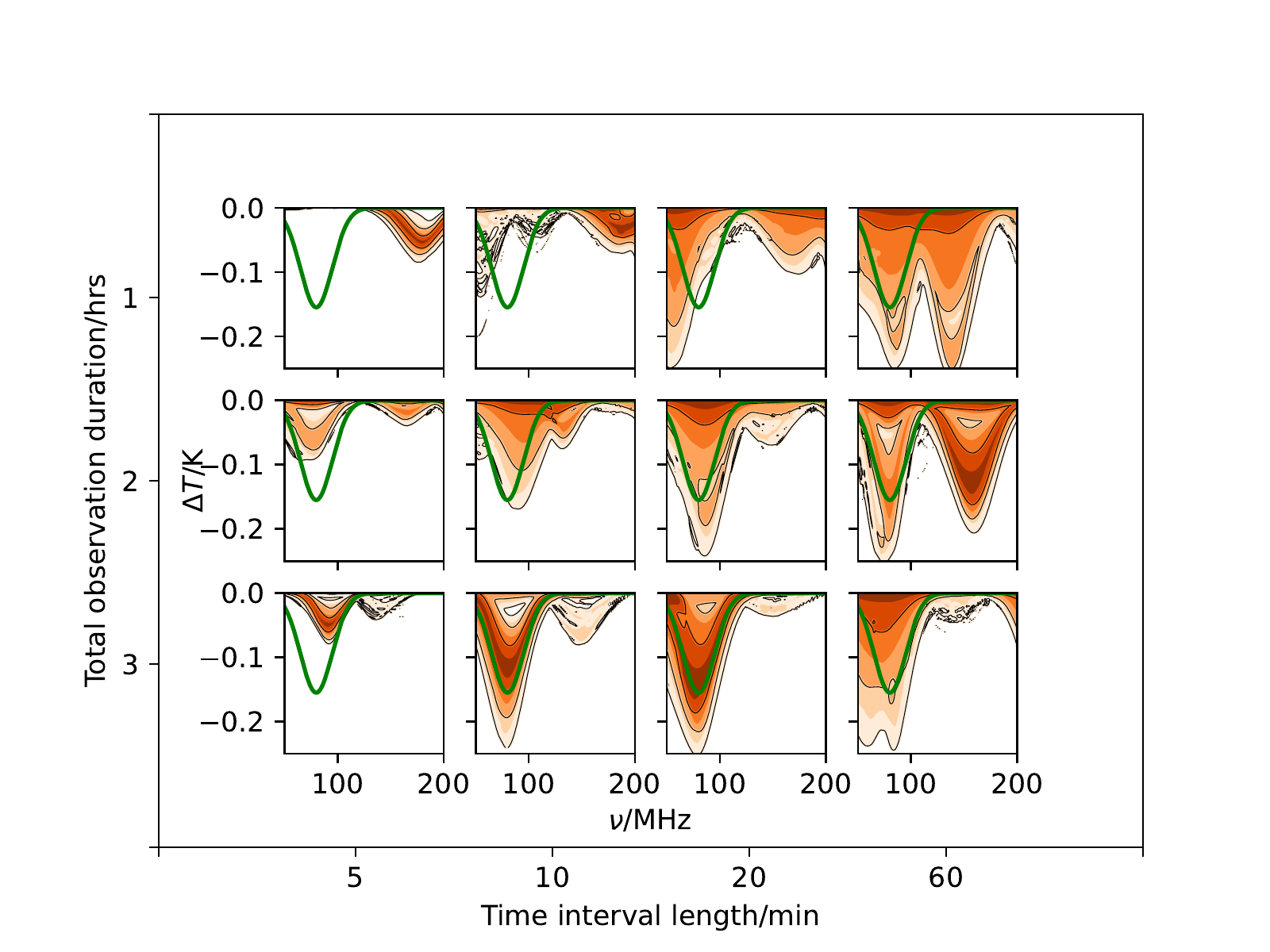}
\caption{Optimum evidence recovered signals for time averaged fitting}\label{fig:cont_op_av_only_hex}
\end{subfigure}
\begin{subfigure}[b]{\columnwidth}
\includegraphics[width=\columnwidth]{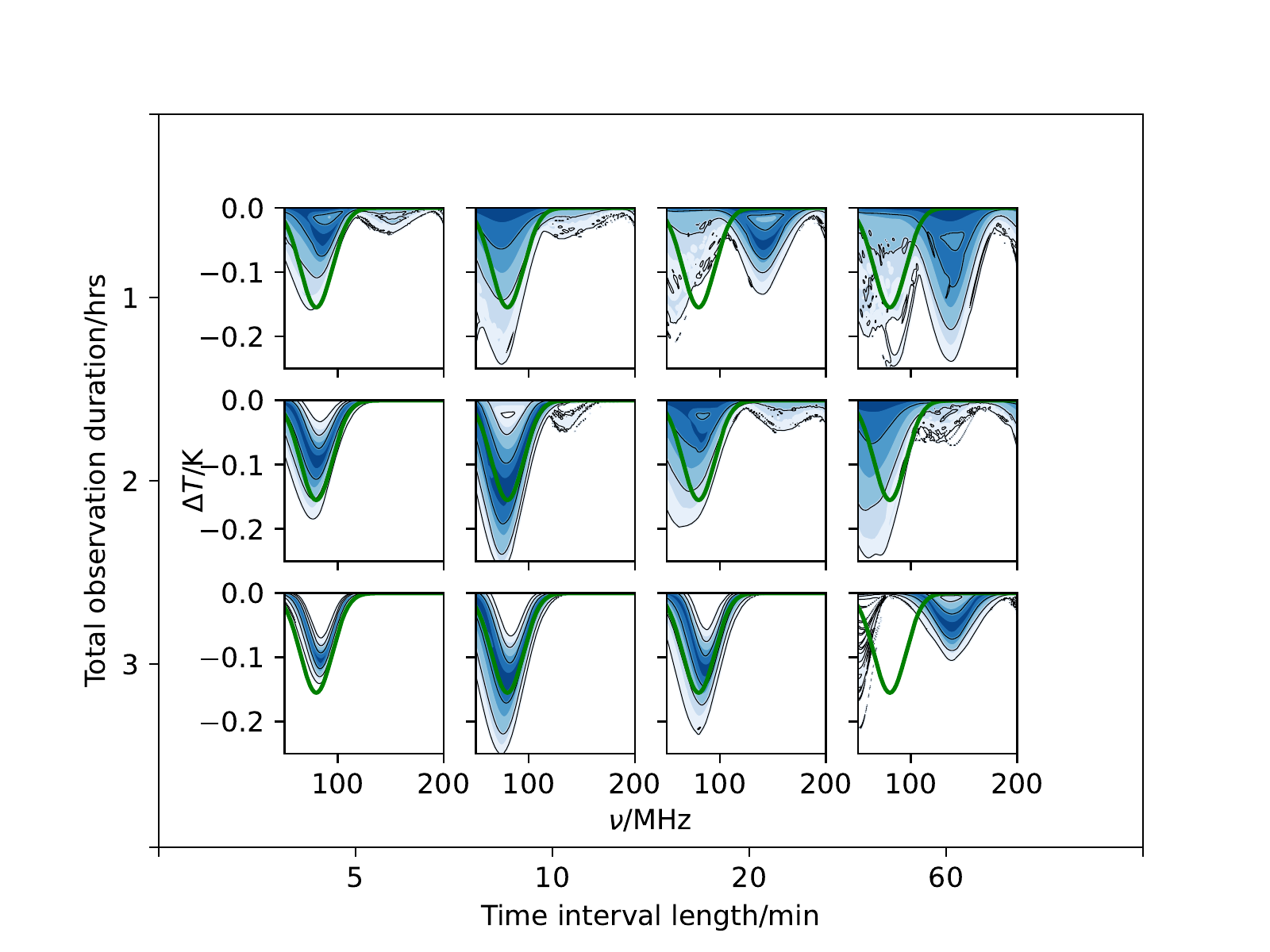}
\caption{Optimum evidence recovered signals for time separated fitting}\label{fig:cont_op_sep_only_hex}
\end{subfigure}
\begin{subfigure}[b]{\columnwidth}
\includegraphics[width=\columnwidth]{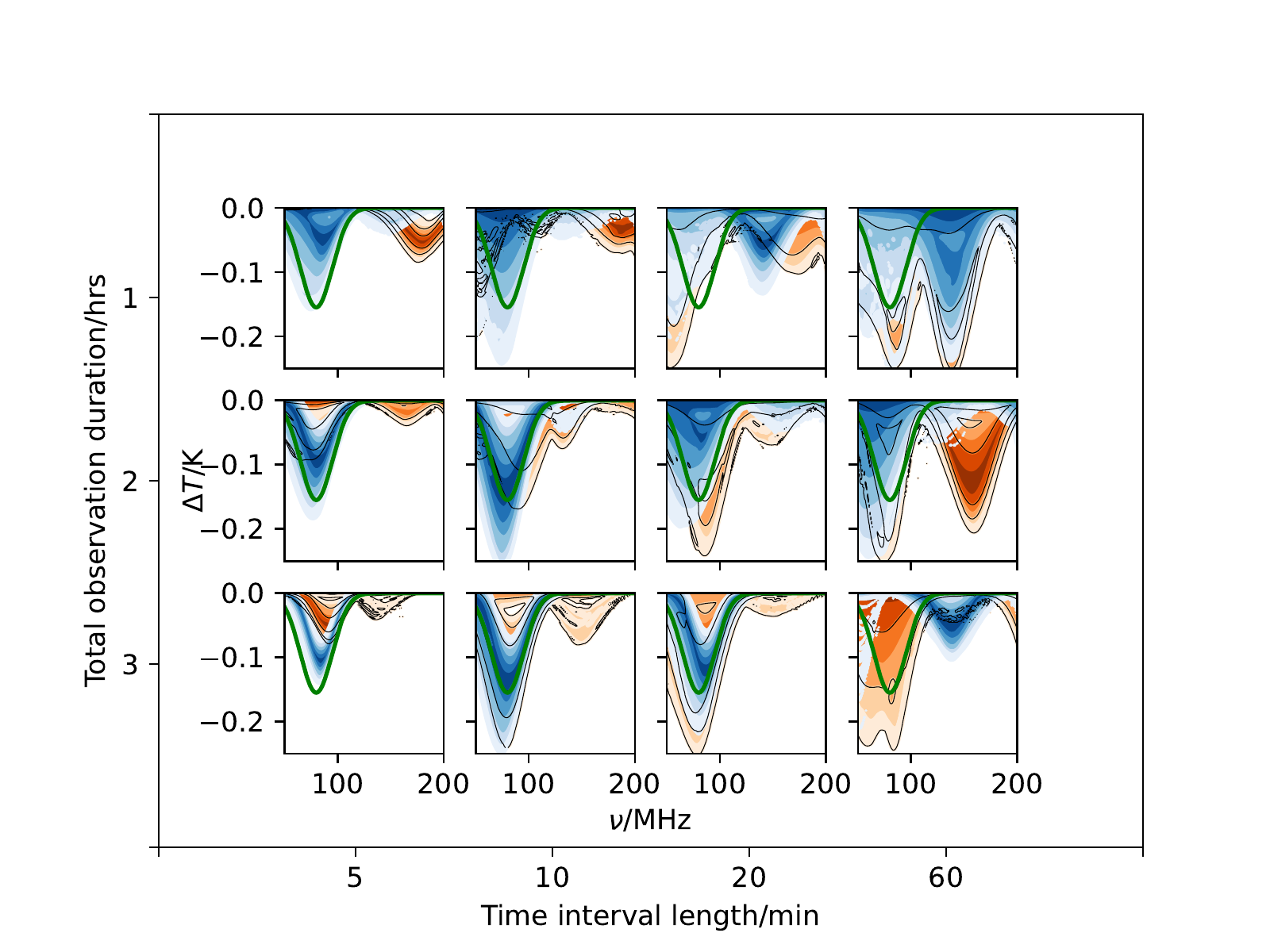}
\caption{Overlay of optimum evidence recovered signals in both cases}\label{fig:cont_op_both_hex}
\end{subfigure}
\begin{subfigure}[b]{\columnwidth}
\includegraphics[width=\columnwidth]{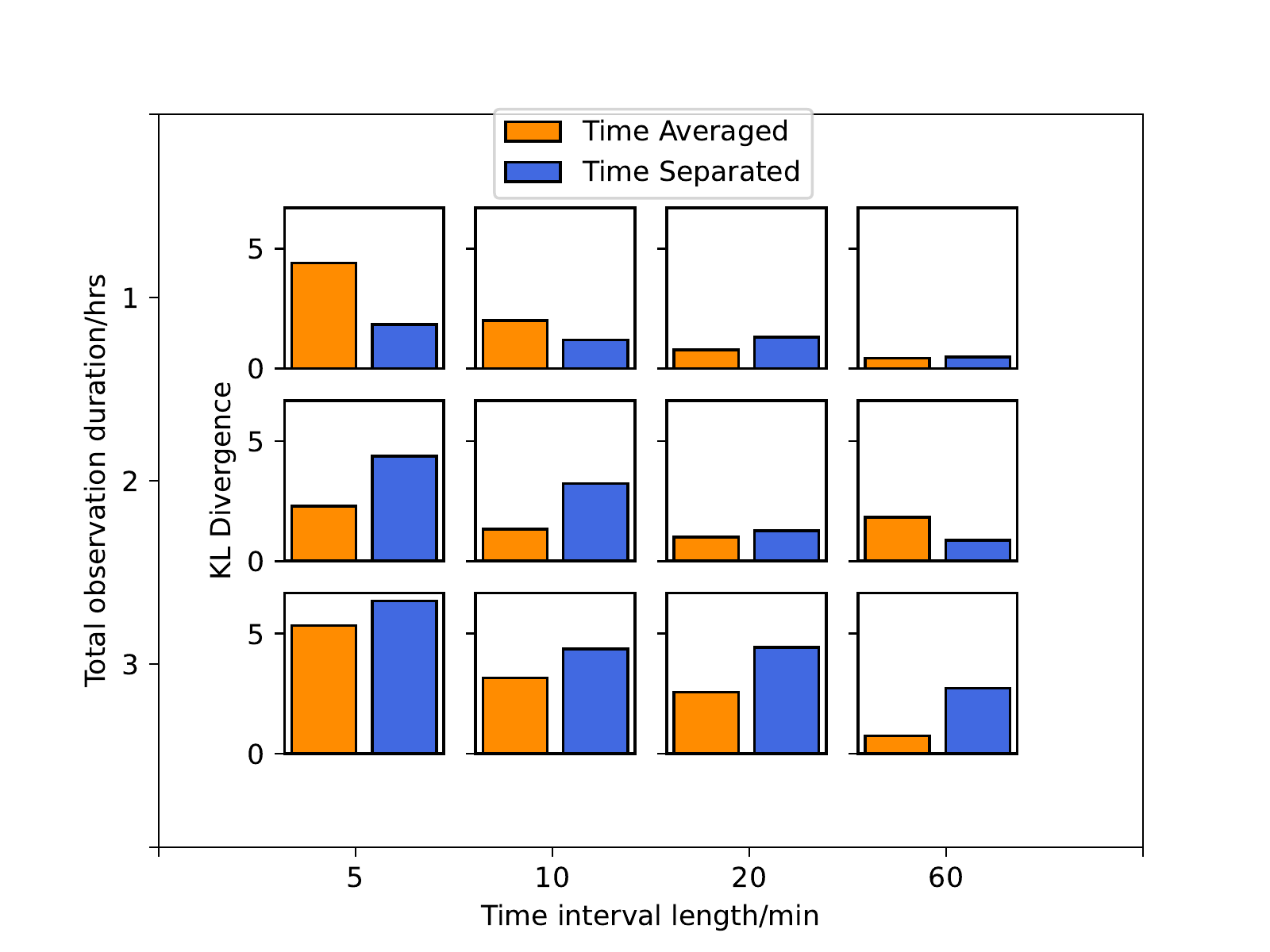}
\caption{Kullback–Leibler divergences of recovered signals}\label{fig:cont_op_KLD_comp_hex}
\end{subfigure}
\caption{Plots of the signal models fit to each simulated data set generated using a hexagonal-bladed dipole antenna beam. The upper plots show the recovered signals using the time averaged likelihood shown in \Cref{eq:integ_L_exp} in orange on the left and using the time separated likelihood, \Cref{eq:sep_L} in blue on the right. Each plot shows the optimum signal for each observation duration and time bin interval tested, and shows the signal for the number of foreground regions $N$ that gave the maximum Bayesian evidence, as shown in \Cref{fig:comp_peak_regions_hex}. In all cases the 'true' signal inserted into the simulated data is shown by a green line, and each colour band corresponds to $\frac{1}{2}\sigma$ uncertainty on the signal model. The bottom left plot shows and overlay of the upper two plots. The bottom right plot shows the Kullback–Leibler divergences of the three signal parameters (centre frequency, width and amplitude) for both fitting methods on each data set.}
\label{fig:cont_signal_results_hex}
\end{figure*}

The recovered signal models for the optimum $N$s of each fitting method and each data set, together with the marginal DKLs of the signal parameters, are shown in \Cref{fig:cont_signal_results_hex}. In the case of 1 hour integration, with 60 minute bins, which corresponds to a single data bin, the two methods of foreground fitting are entirely identical by definition. Correspondingly, it can be seen that in this case, the two methods require the same number of foreground parameters and give almost identical signal reconstructions, with almost equal DKLs, as expected. This signal recovery is, however, quite poor, with very large errors and a very low DKL showing little information content in the signal parameters. Similarly, poor signal recovery can also be seen for all data sets with the shortest integration times and fewest time bins. 

For both the time-averaged and separated cases, the signal recovery can be seen to become more accurate to the true signal as the number of time bins used in the data set increases. In particular, the recovered signals become more accurate as the duration of the observation increases from 1 hour to 3 hours. However, the results of the time-separated modelling process show a much more significant improvement in recovered signal accuracy with both increasing integration time and shorter intervals between successive time bins. As a result, for the longest integration times and finest time binnings, the signals recovered from the data by the time separated modelling method are both significantly more accurate and more precise, with lower error ranges, than those recovered from the same simulated data set by time-averaged models. In several cases, such as with 5 and 10 minute divisions with 2 hours of total integration, this results in the time-separated fitting process successfully recovering an accurate 21cm signal where the time-averaged fit failed to recover any signal, giving a signal reconstruction centred on zero.

These results are corroborated by the KDEs. For both fitting methods, the DKLs, in general, are seen to increase with both increasing integration time and decreasing interval time, in agreement with the improving accuracy of the signal recovery. Furthermore, the DKLs of the parameters recovered by the time-separated method increase to a greater degree than those of the time-integrated fitting. As a result, at the longest integration times and shortest intervals, the time-separated DKLs are consistently higher than the time-averaged. Again, this is in agreement with the observed signal recoveries.

There are a few exceptions to this, however, in which the time-averaged signal has a higher DKL than the time-separated. These are the minority of cases, with only 1 hour integration with 5 and 10 minute divisions, and 2 hours with 60 minute divisions, showing this. These are all observed to occur at either short integration times or very course time binning. In addition, it can be seen that in all three of these cases, the time-separated method still shows a more accurate signal recovery, with the true signal falling in lower sigma regions of the recovered signal distribution than that of the corresponding time-averaged fit.

Overall, therefore, these results demonstrate two main concepts. Firstly, they demonstrate the effect that data that more thoroughly samples the foreground, particularly by observing for a longer period, enables a more accurate signal recovery. This is an expected effect, given that additional data bins reduce the impact of noise, either by averaging it to a lower level in the case of the time averaged modelling, or by informing the model more thoroughly with more data sets in the case of the time separated model. However, the more thorough sampling of the foregrounds could also contribute to this effect, in combination with the method for modelling foregrounds and antenna chromaticity in global 21cm experiments described in A21. 

Secondly, it is shown that the method proposed here, in which each time bin of the data is simultaneously fit to a corresponding foreground model, all contributing to the same parameter fit, results in significant improvement in the accuracy with which the 21cm signal is recovered, relative to simply fitting a single model to time averaged data. This improvement in signal recovery is also seen to become more pronounced the more thoroughly the foregrounds are sampled. As the time separated and time average model fit to the same data sets with the same noise, this improvement, therefore, must arise from improved modelling of the foregrounds, demonstrating the two-fold benefit of additional time bins in the data, which both reduces noise and provided additional data about the foreground structure to enable more accurate foreground modelling.

In addition to the signal recovery, therefore, the foreground modelling can also be considered. \Cref{fig:comp_peak_regions_hex} shows that, in general, the time separated fitting method has optima at higher numbers of foreground parameters than the time averaged method. This shows that the time separated method statistically favours more complex and detailed foreground models to accurately fit the data. Apart from the case where the two fitting models are entirely equivalent, which shows they require equal numbers of foreground parameters, as expected, there are only 2 cases seen of the 12 tested in which the time averaged method reaches an evidence maximum at a higher number of foreground parameters than the time separated method. It should also be noted that, as was shown in \Cref{fig:cont_signal_results_hex}, this increase in the number of foreground parameters used in the model fit does not result in any additional uncertainty in the signal recovery. Instead the signal recovery was seen to improve. This can be attributed to the time separated data, due to not averaging out the changes in the foregrounds with the rotation of the Earth, containing additional information about the nature of the foregrounds, and thus enables the constraint of a greater number of parameters.

\begin{figure*}
\centering
\begin{subfigure}[b]{0.8\textwidth}
\includegraphics[width=\textwidth]{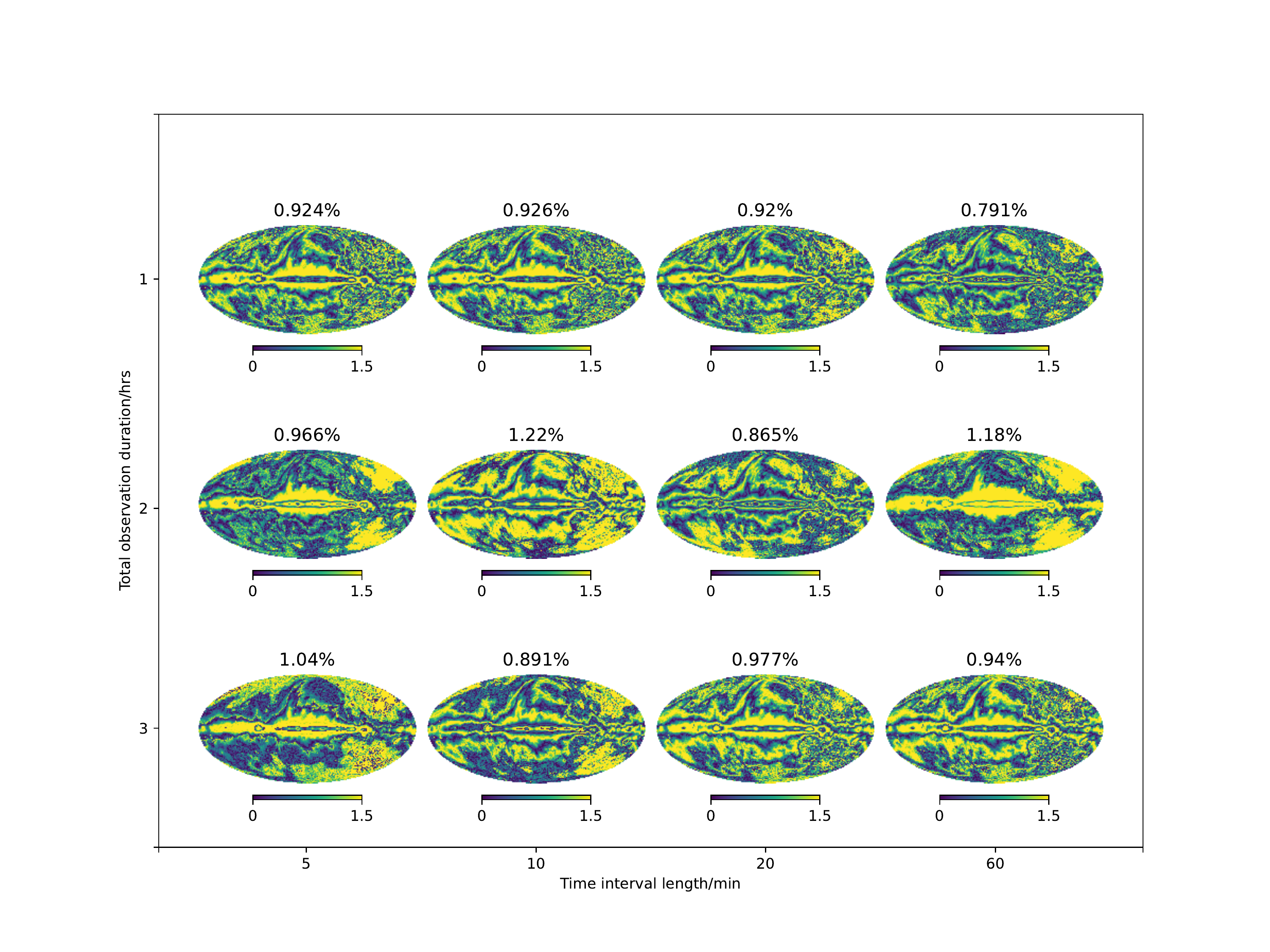}
\caption{Percentage difference between the true and fitted spectral index maps for time averaged modelling.}\label{fig:cont_fore_diff_av_hex}
\end{subfigure}
\begin{subfigure}[b]{0.8\textwidth}
\includegraphics[width=\textwidth]{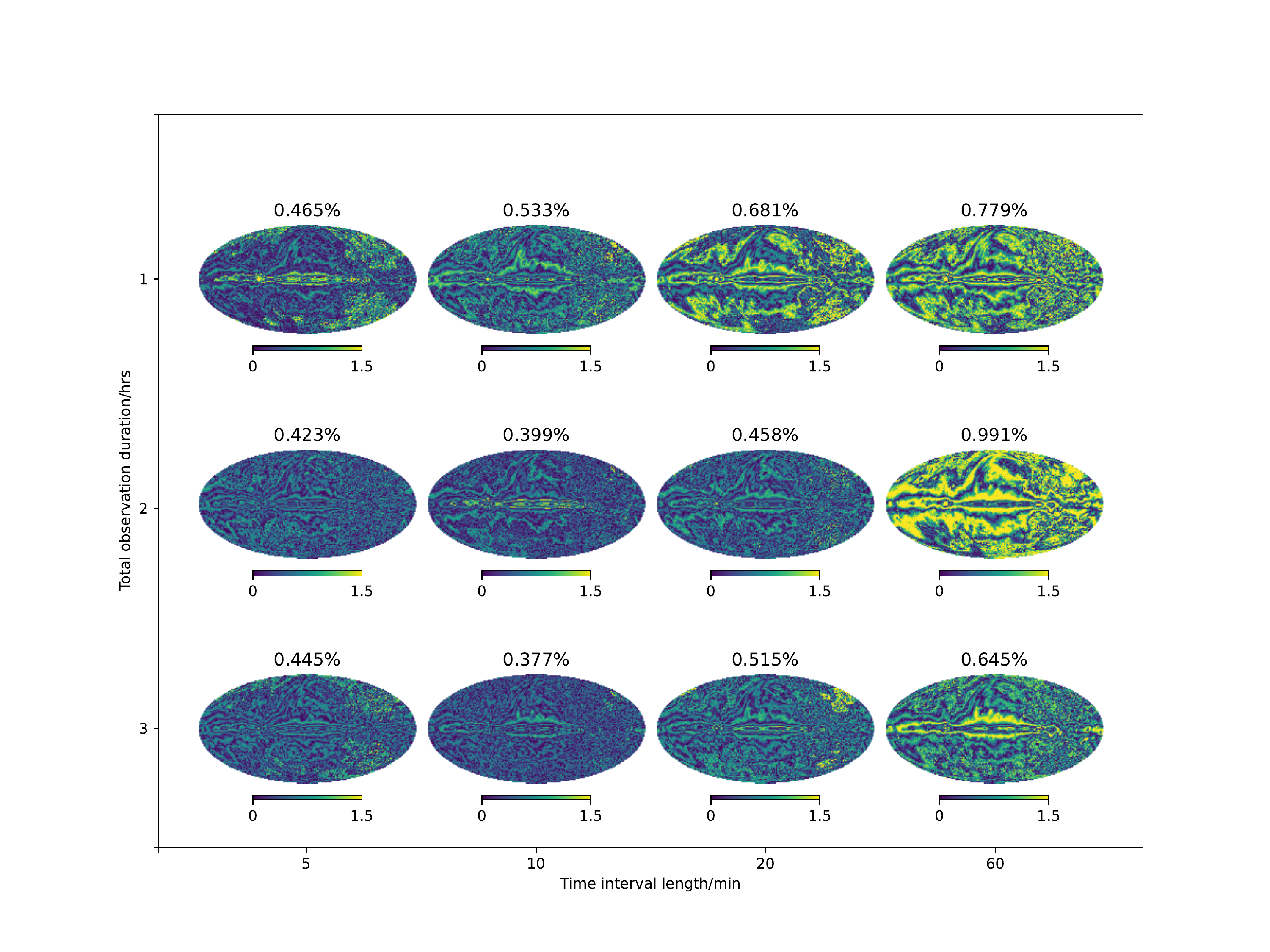}
\caption{Percentage difference between the true and fitted spectral index maps for time separated modelling.}\label{fig:cont_fore_diff_sep_hex}
\end{subfigure}
\caption{Plots of the absolute percentage difference between the `true' spectral index map $\beta\left(\Omega\right)$ used to generate simulated data according to \Cref{eq:sim_data_gen} and the best fit spectral index map for each hexagonal dipole antenna data set. The top plot shows the results for time averaged fitting and the bottom plot show time separated fitting. The average percentage differences are recorded above each plot.}
\label{fig:cont_fore_diff_maps_hex}
\end{figure*}

This increase in model complexity with no drop in signal recovery implies the time-separated method is fitting the foregrounds more accurately. In order to quantify the accuracy of the fitted foregrounds, we compare the fitted spectral index parameters to the `true' spectral index map defined in \Cref{eq:sim_data_gen} that was used to generate the data. This is achieved by finding the reconstructed spectral index map by assigning the weighted posterior average spectral index value of each region to that region, then calculating the absolute percentage difference between each pixel of this map and the `true' $\beta\left(\Omega\right)$ map. \Cref{fig:cont_fore_diff_maps_hex} shows these difference maps for the optimum $N$ case of each data set and fitting method.

It can be seen from \Cref{fig:cont_fore_diff_maps_hex} that, for the time averaged fitting method, the accuracy with which the foreground spectral index parameters are recovered is approximately constant for all data sets tested. The reconstructed spectral index maps are seen to be accurate to the `true' map used to generate the simulated data to approximately 1\% in all tested cases.

For the time separated fitting method, however, the accuracy of the reconstructed foreground model shows a trend of increase with increasing observation duration and finer time binning. This is equivalent to the improvement that was seen in the recovered signal. For the data set of 1 hour observation with 1 hour between bins, which is a data set with only 1 time bin, the two methods have a very similar foreground accuracy, as expected. However, in every other tested case, the time separated model fitting gives a reconstructed foreground spectral index map that is significantly more accurate to the `true' map that the time averaged case. This reaches improvements relative to the time averaged case of factors of 2-3 for longer observation time and shorter intervals.

These results demonstrate that jointly fitting separated data time bins in the manner described here significantly improves the accuracy with which radio foregrounds can be modelled, in addition to improving 21cm signal recovery. Therefore, this process, in combination with a model that parameterises the foregrounds based on a physical property, such as that of A21, improves the ability of a Global 21cm experiment to study the radio foregrounds, in addition to the intended 21cm cosmology.

\subsection{Logarithmic Spiral}\label{comparison_log_spiral}
The previous analysis was performed for a hexagonal-bladed dipole antenna. This is one of the antennae that will be used by the experiment REACH, and is similar to the rectangular bladed dipole used in EDGES \citep{bowman18}. However, \citet{anstey22} showed that the modelling process presented in A21 performs differently for different antenna designs. Therefore, in order to investigate how the performance of time-separated fitting is affected by the antenna being used, this analysis was repeated for a different antenna. 

For this purpose, a logarithmic spiral antenna was used, as it is highly distinct from a dipole antenna and is also intended to be used in REACH. \citet{anstey22} found that the A21 modelling process was able to recover global 21cm signals in data from a log spiral antenna more accurately and more reliably than in that of a dipole, owing to the beam having smaller and more regular chromatic variations than a dipole.

The analysis was therefore repeated with this antenna. 12 simulated data sets were generated according to \Cref{eq:sim_data_gen}, taking $D\left(\Omega,\nu\right)$ as a log spiral antenna beam, according to the time bin layouts shown in \Cref{fig:comp_bins}. Mock Gaussian 21cm signals with a centre frequency of 80MHz, a width of 15MHz and an amplitude of 0.155K were injected to every time bin of each data set. The resulting data sets were hen jointly fit with a foreground model using the A21 method and a signal model, using both time averaged fitting and time separated fitting, for a wide range of $N$s.  

\begin{figure}
    \centering
    \includegraphics[width=\columnwidth]{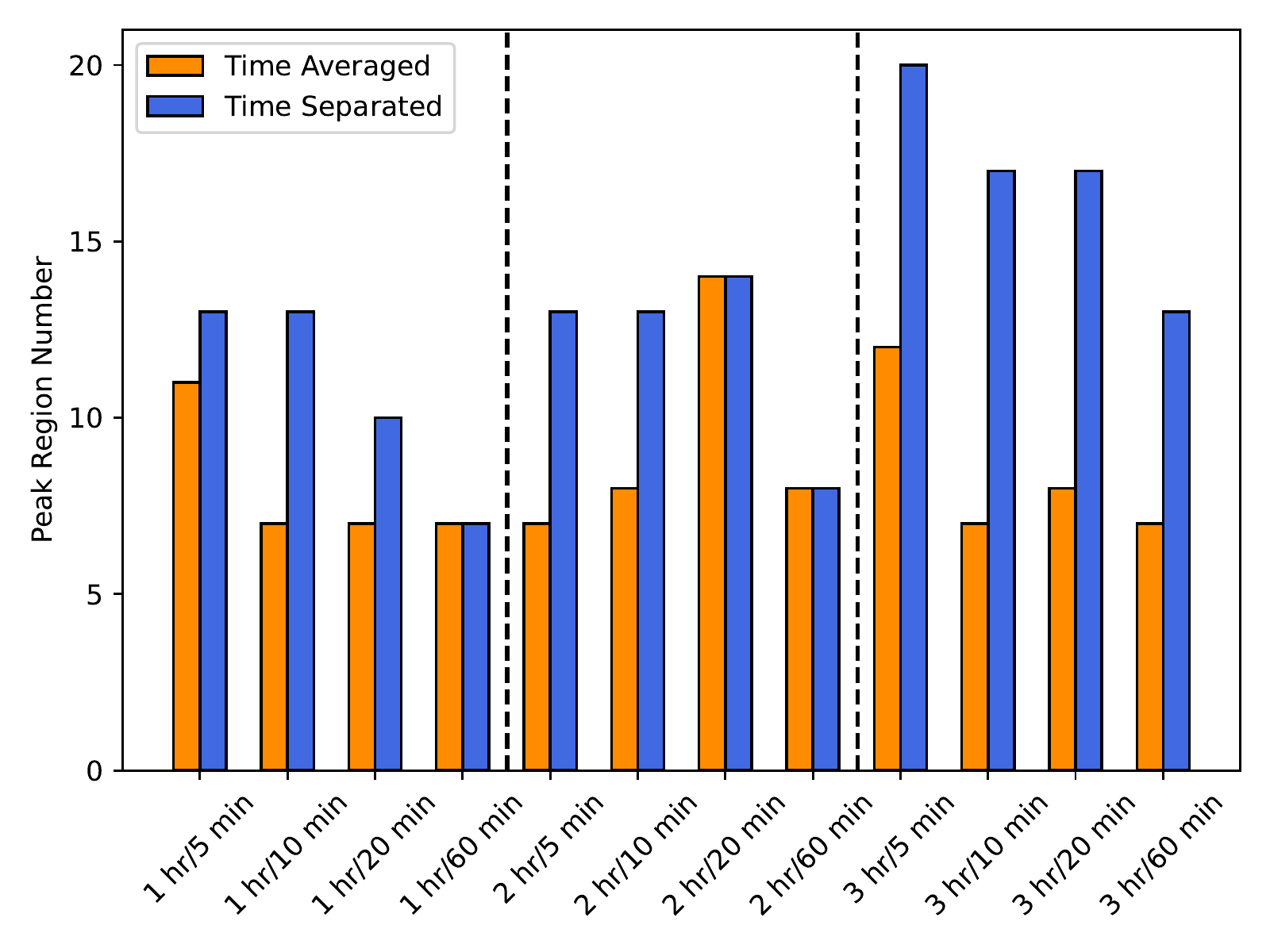}
    \caption{Plot of the number of regions $N$ that the foreground model was divided into in order to give the highest evidence model fit for each of the simulated data sets shown in \Cref{fig:comp_bins}, generated using a log spiral antenna.}
    \label{fig:comp_peak_regions_ls}
\end{figure}

\begin{figure*}
\centering
\begin{subfigure}[b]{\columnwidth}
\includegraphics[width=\columnwidth]{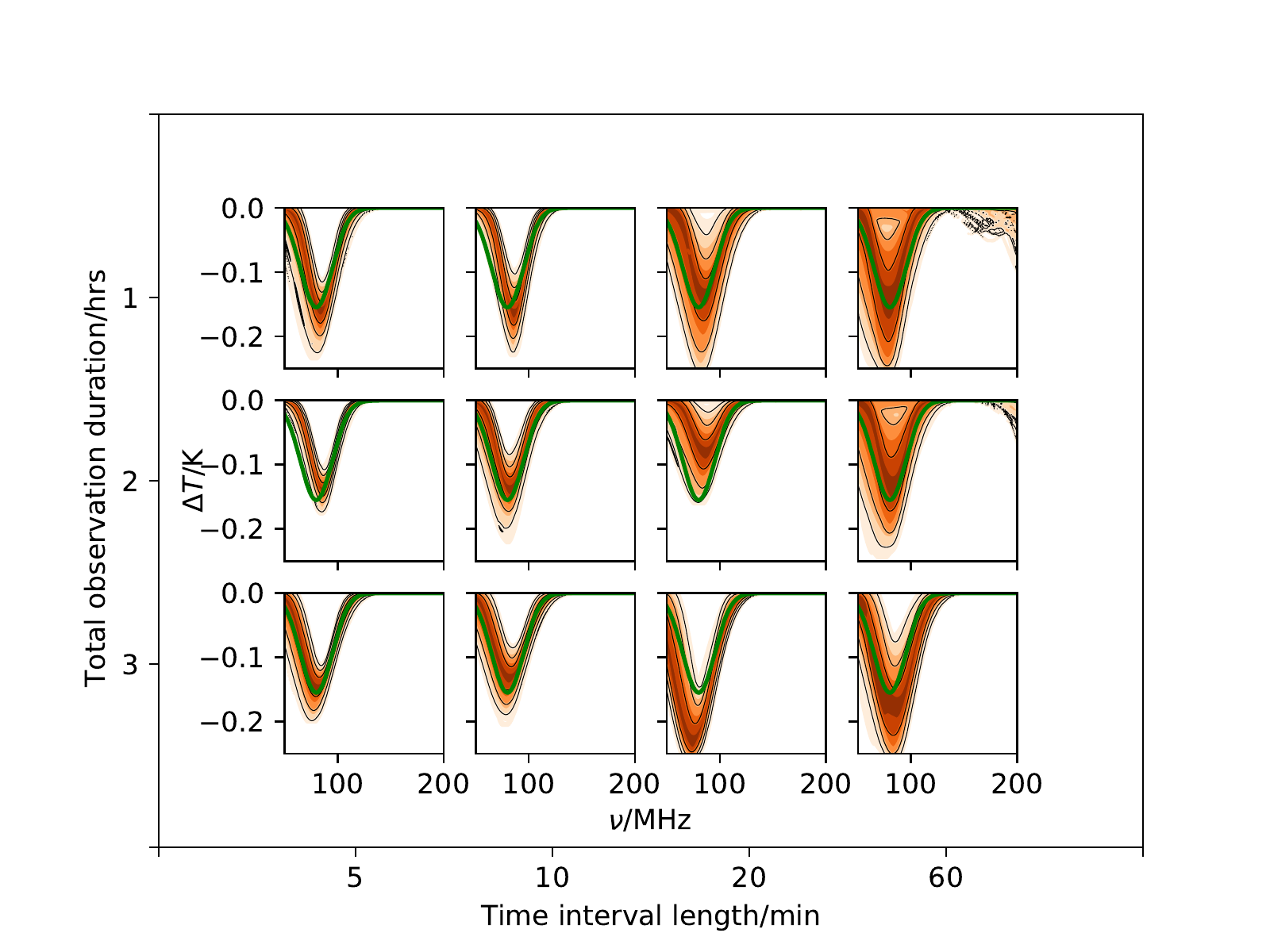}
\caption{Optimum evidence recovered signals for time averaged fitting}\label{fig:cont_op_av_only_ls}
\end{subfigure}
\begin{subfigure}[b]{\columnwidth}
\includegraphics[width=\columnwidth]{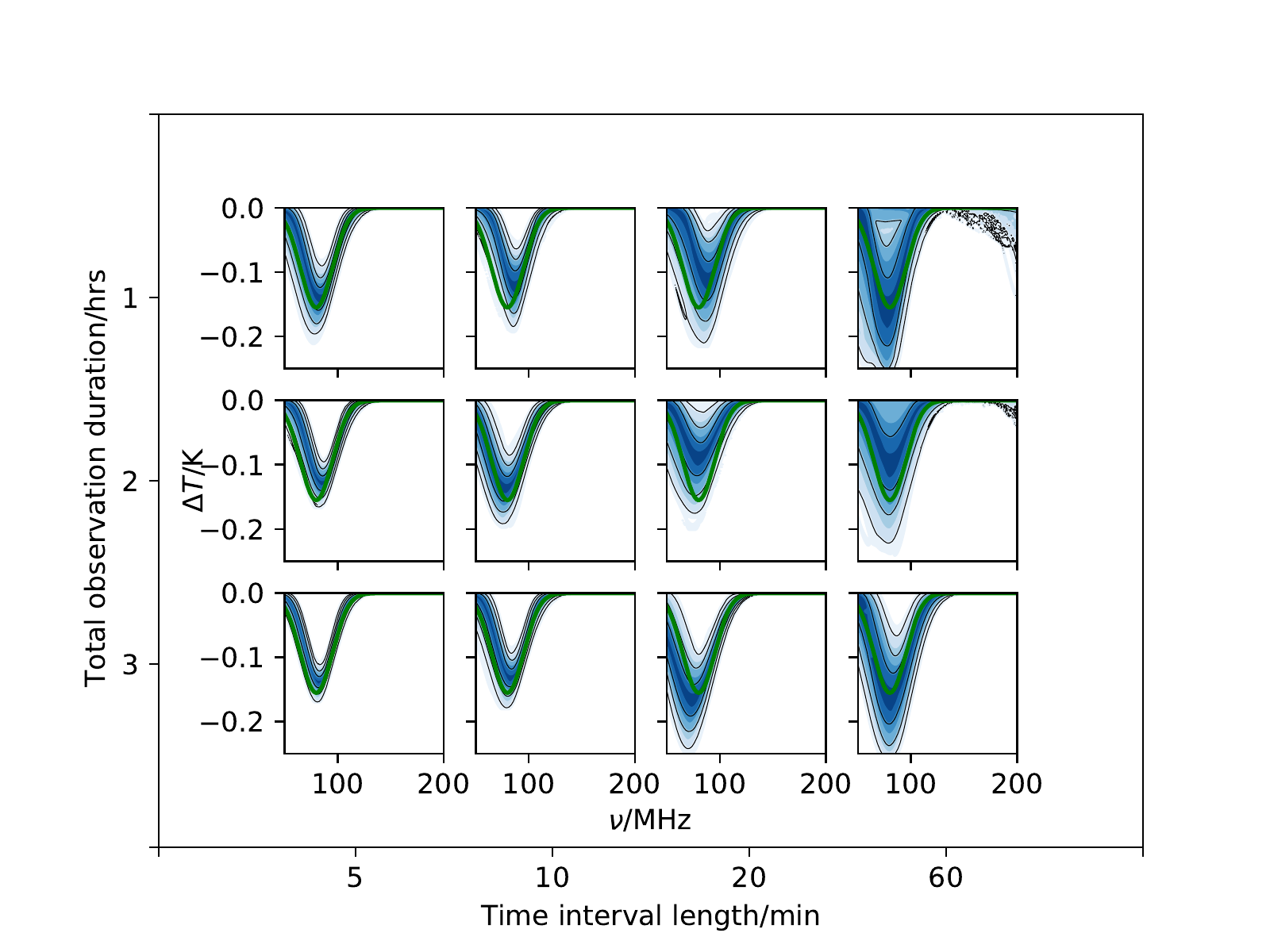}
\caption{Optimum evidence recovered signals for time separated fitting}\label{fig:cont_op_sep_only_ls}
\end{subfigure}
\begin{subfigure}[b]{\columnwidth}
\includegraphics[width=\columnwidth]{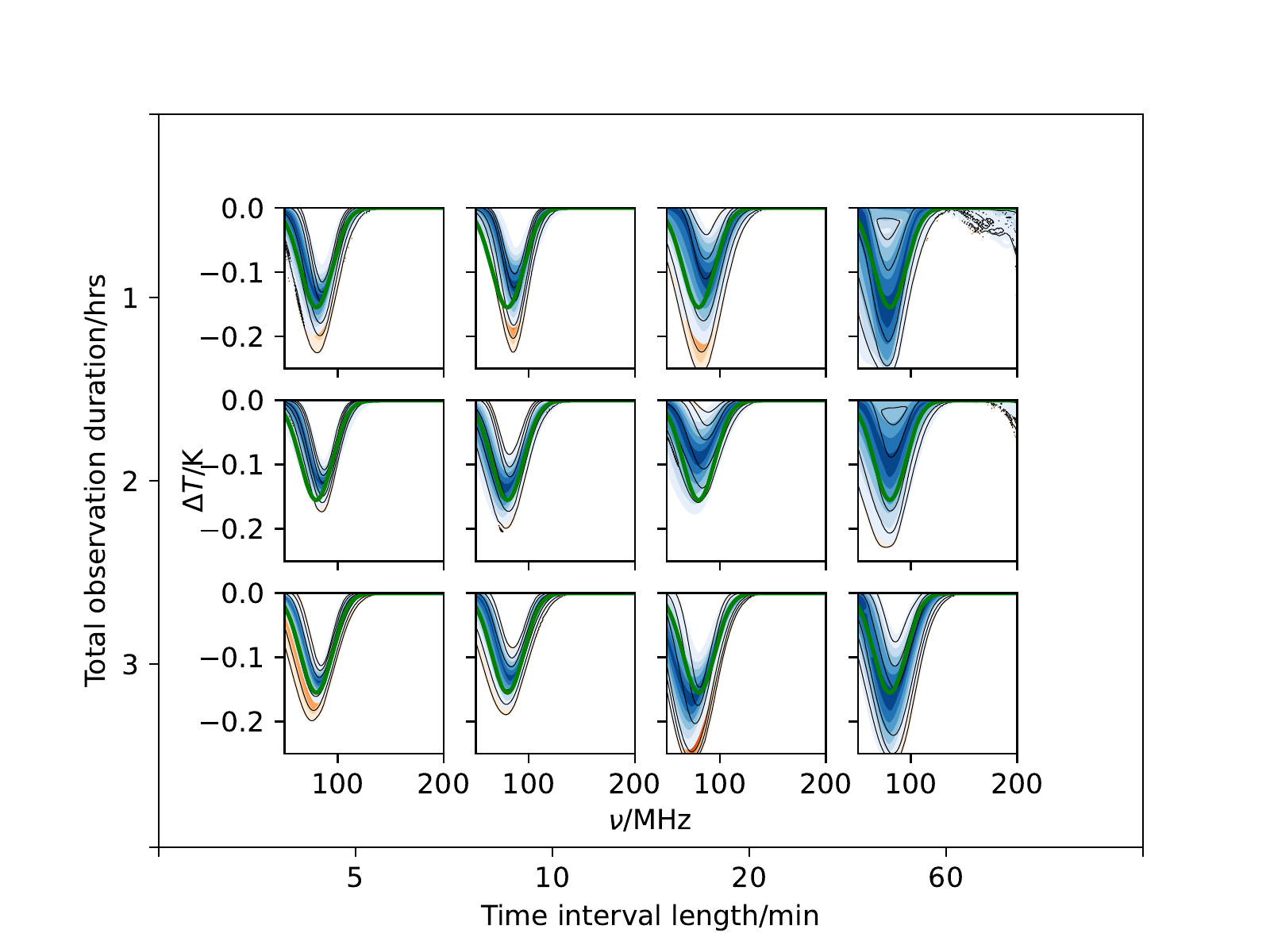}
\caption{Overlay of optimum evidence recovered signals in both cases}\label{fig:cont_op_both_ls}
\end{subfigure}
\begin{subfigure}[b]{\columnwidth}
\includegraphics[width=\columnwidth]{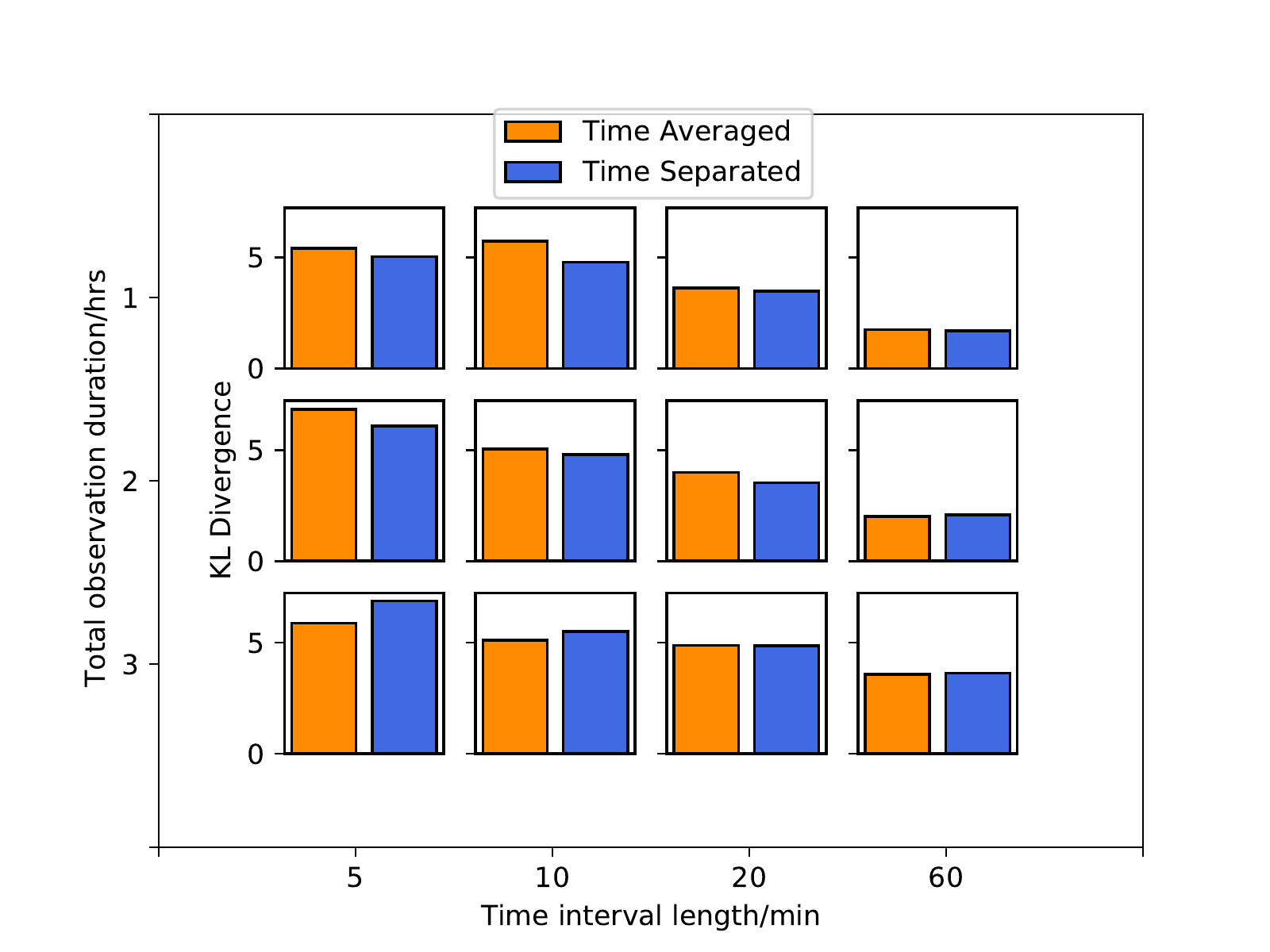}
\caption{Kullback–Leibler divergences of recovered signals}\label{fig:cont_op_KLD_comp_ls}
\end{subfigure}
\caption{Plots of the signal models fit to each simulated data set generated using a log spiral antenna beam. The upper plots show the recovered signals using the time averaged likelihood shown in \Cref{eq:integ_L_exp} in orange on the left and using the time separated likelihood, \Cref{eq:sep_L} in blue on the right. Each plot shows the optimum signal for each observation duration and time bin interval tested, and shows the signal for the number of foreground regions $N$ that gave the maximum Bayesian evidence, as shown in \Cref{fig:comp_peak_regions_ls}. In all cases the 'true' signal inserted into the simulated data is shown by a green line, and each colour band corresponds to $\frac{1}{2}\sigma$ uncertainty on the signal model. The bottom left plot shows and overlay of the upper two plots. The bottom right plot shows the Kullback–Leibler divergences of the three signal parameters (centre frequency, width and amplitude) for both fitting methods on each data set.}
\label{fig:cont_signal_results_ls}
\end{figure*}

The number of foreground regions required to give the highest evidence fits in each case are compiled in \Cref{fig:comp_peak_regions_ls}. The signal fits for these optimal $N$ cases, and the corresponding DKLs, are shown in \Cref{fig:cont_signal_results_ls}.

Firstly, these results show more accurate signal recovery in all cases tested than for the hexagonal dipole antenna, as is expected from the results of \citet{anstey22}. These results also show again a very clear trend of improving accuracy of signal recovery with both increasing observation time and decreasing time intervals. However, unlike for the hexagonal dipole, this effect occurs comparably for both modelling methods. As a result, there is no clear difference in the recovered signals between the two methods for any of the tested cases.

These results are supported by the DKLs of the recovered signals. Both fitting methods show clear trends of increasing DKL with increasing observation time and decreasing intervals. However, the changes are comparable for both modelling methods and there are no clear trends in the differences between the two. This shows that, for an antenna with a more regular and easily modelled chromaticity that enables more reliable 21cm signal detection such as a log spiral, performing simultaneous model fits to separated data time bins does not produce any significant improvement in the ability to recover the signal relative to simply fitting a single model to a time-average of the data. This is in contrast to an antenna with a more complex chromatic structure like the hexagonal dipole, where a clear benefit was seen.

This gives some insight into how this simultaneous fitting of time bins aids in 21cm signal recovery. Given that the improvement in signal recovery produced by fitting the data in this manner is significantly larger for an antenna with a less easily modelled chromatic distortion, it suggests that this improvement is due to the chromatic distortion being constrained in the model more precisely. This is reasonable, given that non-averaged data contains additional data about the form of the radio foregrounds based on their change from the Earth's rotation. The chromatic distortions that arise in global 21cm data due to coupling of the antenna chromaticity with these foregrounds are therefore, by definition, strongly coupled to the foregrounds. This suggests that non-averaged data also contains additional information about the form of the chromatic distortion. For observations where the antenna chromaticity is small, regular and easily modelled, this additional information is negligible relative to that already present in the averaged data, and so exploiting it through simultaneous fitting of time binned data sets gives no significant improvement to the signal recovery. However, for an antenna with a less easily modelled chromatic structure, this additional information gained from the change in foregrounds and chromatic distortion with the Earth's rotation is significant, and exploiting it enables the chromaticity to be modelled much more accurately and the global 21cm signal to be detected with correspondingly higher precision. This process of simultaneous fitting, therefore, is highly advantageous to global 21cm experiments.

Although the model fits for simulated log spiral data showed no significant improvement in signal recovery, the foreground fits can also be considered. \Cref{fig:comp_peak_regions_ls} shows that, similarly to the hexagonal dipole case, the time separated modelling process results in the number of foreground parameters require for the Bayesian evidence to peak is, in general, larger than for the time averaged modelling. These results actually show no cases where the time averaged model requires more foreground parameters than the time separated case. There are, however, excluding that where the two processes are identical, 2 cases where the two methods require equal numbers of foreground parameters. This reinforces the conclusion from the hexagonal dipole results, showing that time separated fitting tends to requires more detailed foreground models to accurately fit the data. It can also be noted that, although no improvement in signal precision is seen from using time separated data, the higher parameter foreground models also do not result in any decrease in the signal precision, owing to the additional information gained from exploiting changes from the Earth's rotation. 

\begin{figure*}
\centering
\begin{subfigure}[b]{0.8\textwidth}
\includegraphics[width=\textwidth]{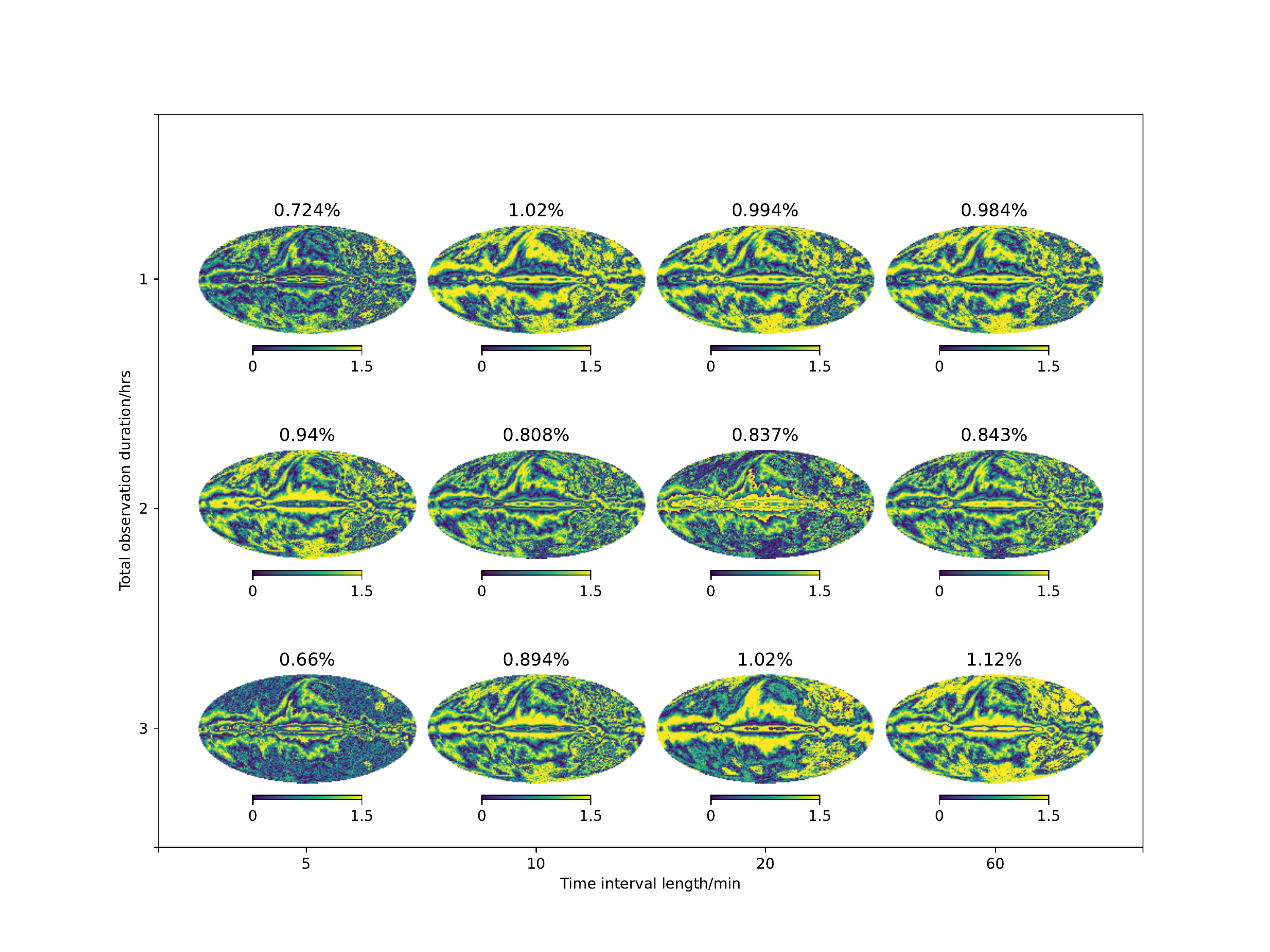}
\caption{Percentage difference between the true and fitted spectral index maps for time averaged modelling.}\label{fig:cont_fore_diff_av_ls}
\end{subfigure}
\begin{subfigure}[b]{0.8\textwidth}
\includegraphics[width=\textwidth]{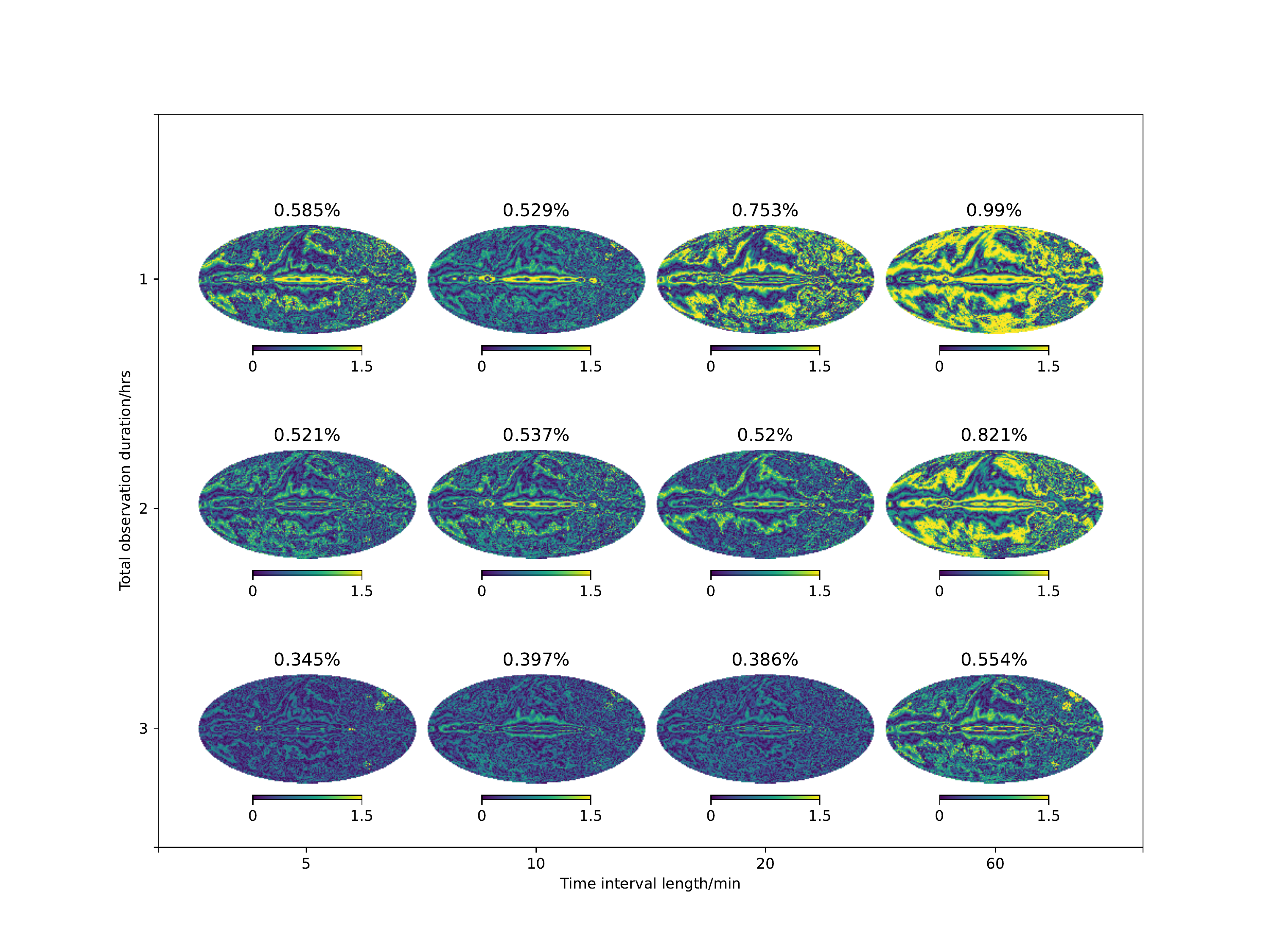}
\caption{Percentage difference between the true and fitted spectral index maps for time separated modelling.}\label{fig:cont_fore_diff_sep_ls}
\end{subfigure}
\caption{Plots of the absolute percentage difference between the `true' spectral index map $\beta\left(\Omega\right)$ used to generate simulated data according to \Cref{eq:sim_data_gen} and the best fit spectral index map for each log spiral antenna data set. The top plot shows the results for time averaged fitting and the bottom plot show time separated fitting. The average percentage differences are recorded above each plot.}
\label{fig:cont_fore_diff_maps_ls}
\end{figure*}

\Cref{fig:cont_fore_diff_maps_ls} shows the absolute percentage differences between the reconstructed maps and the `true' map used to generate the simulated data. Unlike for the hexagonal dipole, these results do show a trend of increasing accuracy of the recovered foreground spectral index map with both increasing observation time and decreasing interval length for the time averaged modelling case. This suggests that the less easily modelled chromaticity of the hexagonal dipole masked the foreground structure to a degree. For the time separated fitting, however, the improvement in foreground accuracy with observation time and finer binning is much more significant, and in line with the effect seen for the hexagonal dipole. As a result, the time separated case again shows significantly more accurate recovery of the foregrounds than the time separated case, by about a factor of 2 for the longest observation times and shortest intervals. This demonstrates that there is still an advantage to using time separated model fitting, even for an antenna with easily modelled chromaticity like the log spiral, as it enables much more accurate and detailed investigation of the radio foregrounds with no cost to the global 21cm signal recovery.

\section{Investigation of variations in data binning}\label{variations}
In the previous section, we demonstrated the benefits to both signal and foreground modelling in 21cm cosmology of utilising changing foregrounds with the Earth's rotation in a joint Bayesian fitting process. However, the process was only tested on data sets with a relatively short observation time. Therefore, in this section, we investigate a wider range of data bin layouts that cover a greater range of LSTs, in order to more completely determine the impact of the LST range covered by the data on signal and foreground recovery. For all the subsequent analysis, we use the log spiral antenna beam.

\subsection{Multiple observation days}\label{variations_fixed_duration}
Firstly, we note that the analysis in the previous section was performed on simulated data of one continuous observation on a single night. However, in practice, global 21cm experiments will observe for many nights, potentially spanning several months. Given that observations at the same clock time on successive nights have a 4 minute difference in sidereal time, each additional night included in the data is theoretically equivalent, in terms of the foregrounds observed, to sampling the hour of LST more finely on a single night.
In order to investigate the impact that extended observations of this kind could have on foreground and signal reconstruction through time separated data modelling, the following simplified analysis was performed.

\begin{figure}
    \centering
    \includegraphics[width=\columnwidth]{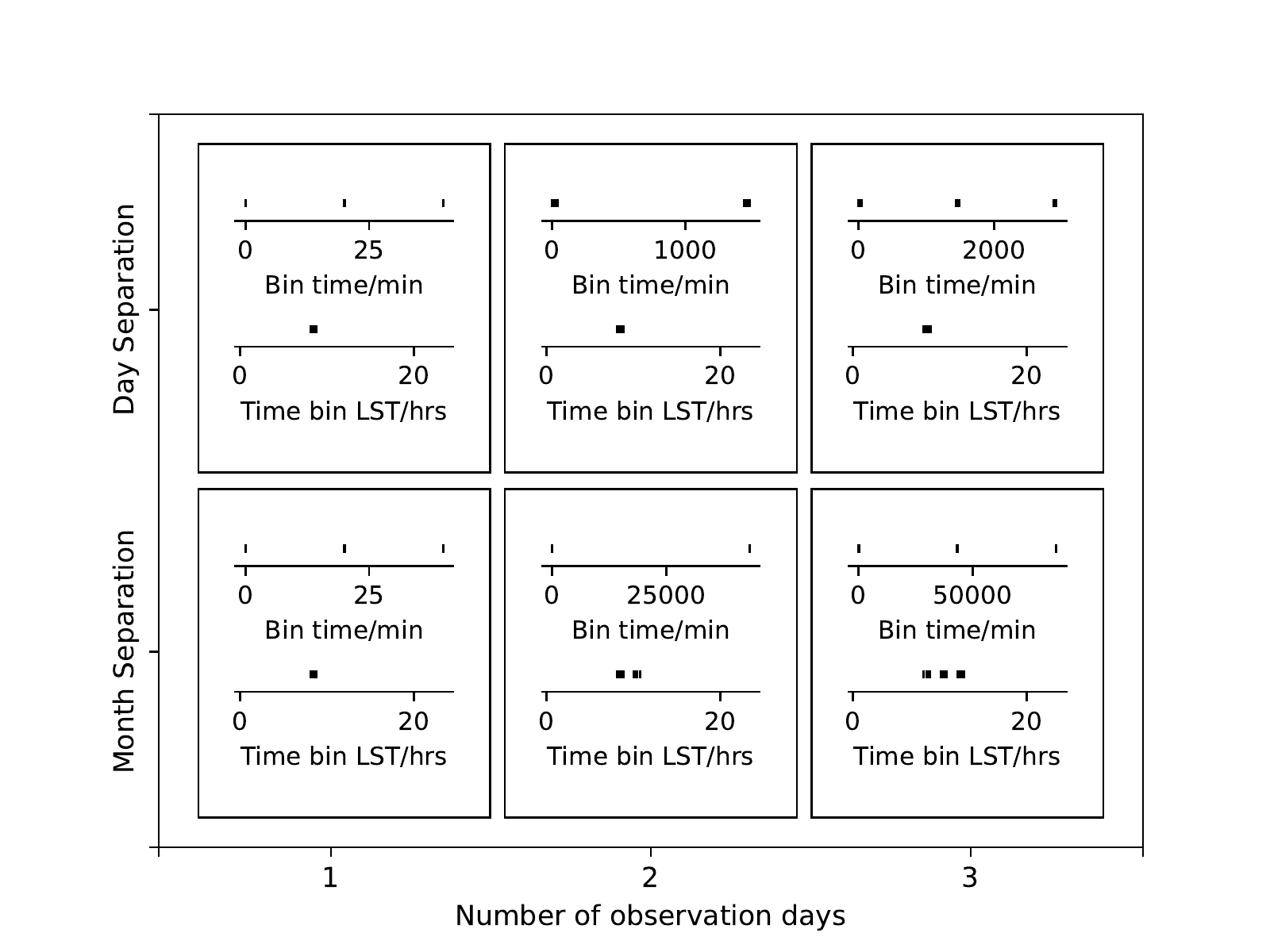}
    \caption{Time bin layouts for tests investigating the effects of observations spanning multiple nights. In each, case a single night of observation consists of one hour of integration with 20 minute between time bins. The number of nights of observations is shown on the x axis and the y axis showing the separation between the nights of observation, with `day separation' referring to 24 hours between each hour of observation and `month separation' referring to 720 hours (30 days) between each hour of observation. In each subplot, the upper plot shows the time of each time bin after the start time of 00:00:00 01-01-2019 UTC and the lower plot shows the LSTs of those time bins.}
    \label{fig:duration_bins}
\end{figure}

Firstly, a new collection of simulated data sets was generated, in which an observation at a fixed clock time was repeated on multiple nights. We took the observation on any single night to be an hour long with 20 minute intervals between bins, giving 3 time bins per night. As in the previous analysis, this assumed an antenna located in the Karoo radio reserve. 6 cases were considered. The first 3 consisted of an hour of observation on a single night, with observations beginning at 00:00:00 01-01-2019, then the same 3 data bins with an additional 3 24 hours later, then those 6, with an additional 3 24 hours later again. In each case, the hour of observation on each night begins at 00:00:00. These cases will be referred to from here as `day separation'. The second set of 3 data sets consist of the same format, but with 720 hours (30 days) between each hour of observation, rather than 24 hours. These will be referred to from here as `month separation'. \Cref{fig:duration_bins} shows the full layout of time bins in these 6 cases.

\begin{figure}
    \centering
    \includegraphics[width=\columnwidth]{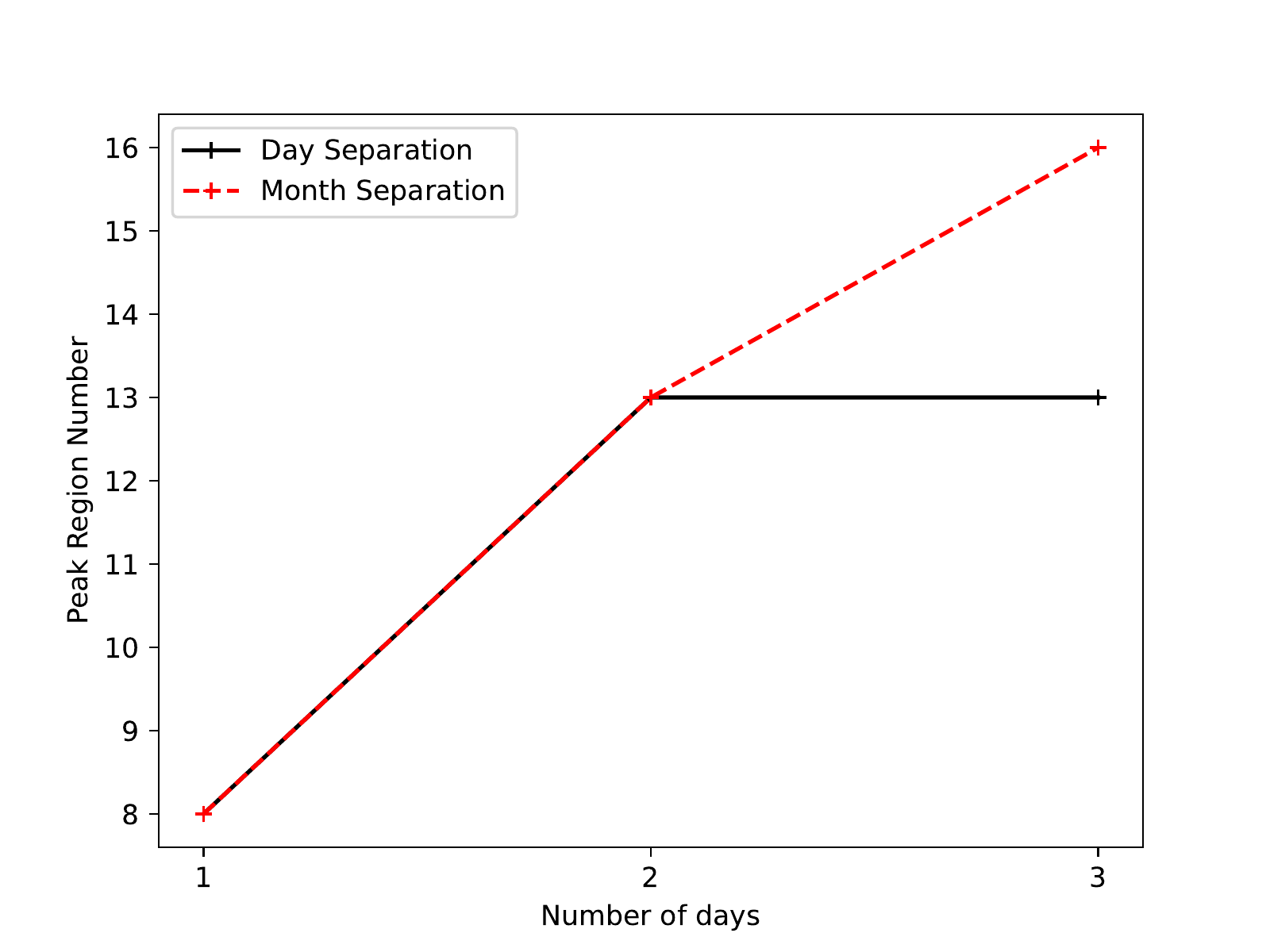}
    \caption{Plot of the number of regions $N$ that the foreground model was divided into in order to give the highest evidence model fit, using the time separated fitting method, for each of the simulated data sets shown in \Cref{fig:duration_bins}, generated using a log spiral antenna.}
    \label{fig:duration_peak_regions}
\end{figure}

Each of these data sets was then fit for, using a range of $N$ in order to find that which gave the optimum Bayesian evidence, via the time separated modelling process. The $N$s that gave the highest Bayesian evidence for each data set are shown in \Cref{fig:duration_peak_regions}.

\begin{figure*}
\centering
\begin{subfigure}[b]{\columnwidth}
\includegraphics[width=\columnwidth]{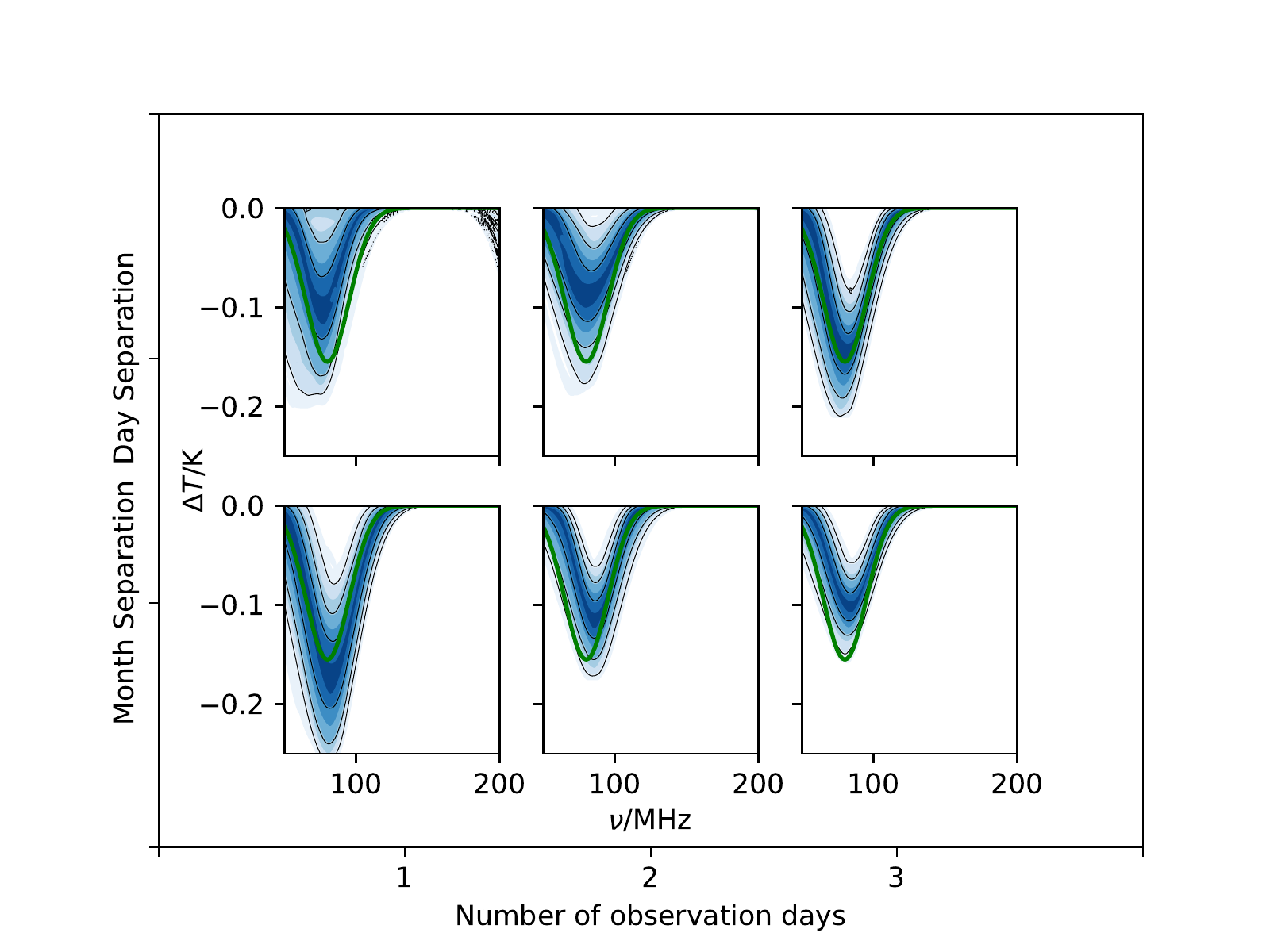}
\caption{Optimum evidence recovered signals}\label{fig:duration_op}
\end{subfigure}
\begin{subfigure}[b]{\columnwidth}
\includegraphics[width=\columnwidth]{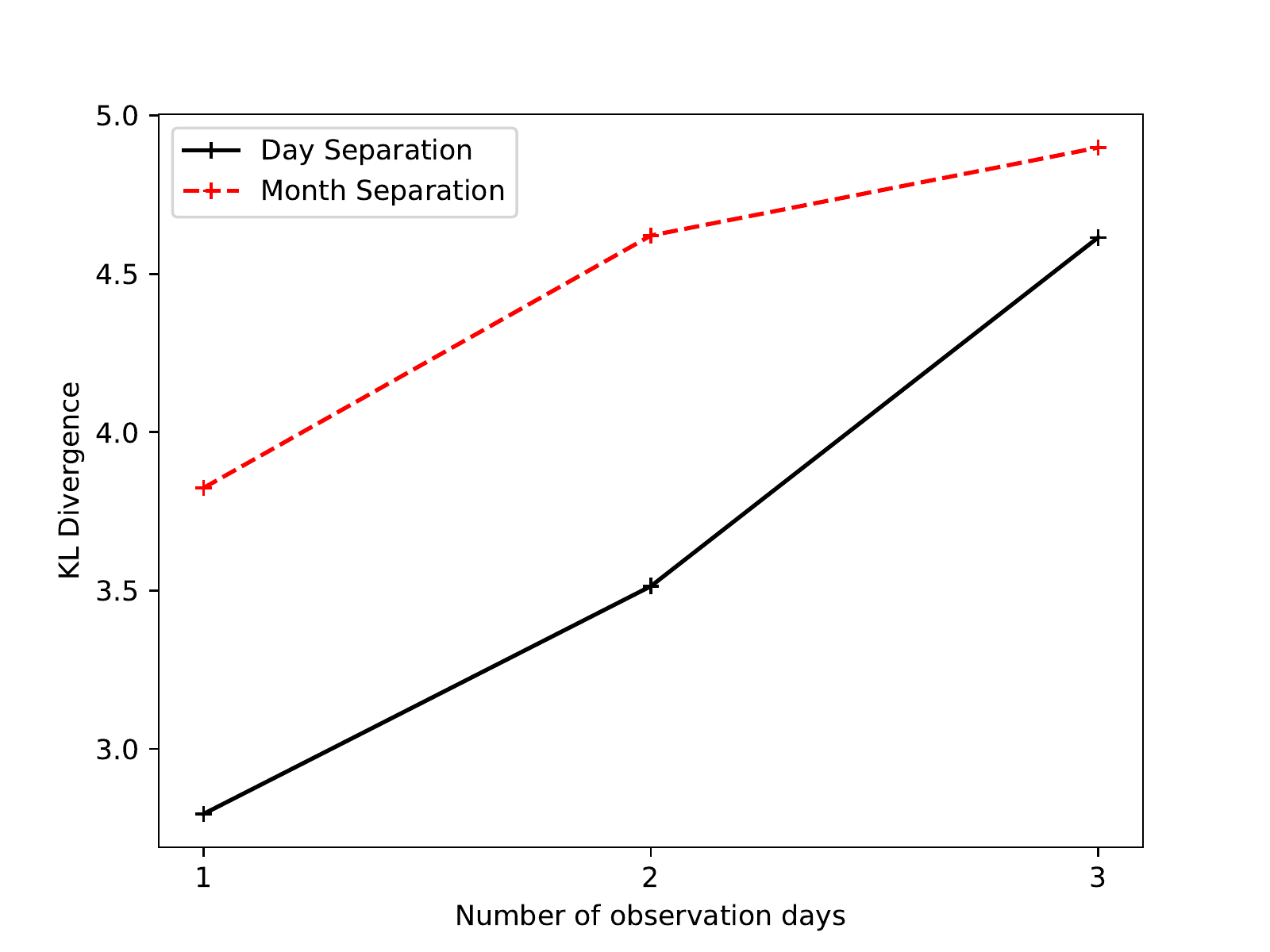}
\caption{Kullback–Leibler divergences of recovered signals}\label{fig:duration_op_KLD}
\end{subfigure}
\caption{Plots of the signal model fits to each simulated data set shown in \Cref{fig:duration_bins}, using the time separated likelihood \Cref{eq:sep_L}, for the number of foreground regions $N$ that gave the maximum Bayesian evidence, as shown in \Cref{fig:duration_peak_regions}. The left plots show the recovered signal model in each case in blue, with each colour band corresponding to a $\frac{1}{2}\sigma$ uncertainty on the signal model. The 'true' signal inserted into the simulated data is shown by a green line. The right plot shows the Kullback–Leibler divergences of the three signal parameters (centre frequency, width and amplitude) for the model fit to each data set.}
\label{fig:duration_signal_results}
\end{figure*}

\Cref{fig:duration_signal_results} shows the recovered signals and marginalised signal parameter DKLs for each of these optimum $N$ fits.

Considering the day separation case first, it can be seen from \Cref{fig:duration_peak_regions} and \Cref{fig:duration_signal_results} that with each additional day of observation included in the data, the precision of the signal recovery improves, with a corresponding increase of DKL. This is the expected result, given that observations at the same clock time correspond to sampling the hour of LST more finely on a single night, in terms of the foregrounds observed. As was shown in \Cref{fig:cont_signal_results_hex} and \Cref{fig:cont_signal_results_ls}, the finer sampling of the foregrounds and reduced impact of noise that arises from additional time bins in the data, results in a corresponding improvement in the accuracy of the signal recovery. 

The month separation case must also be considered, however. As can be seen in \Cref{fig:duration_bins}, each additional day of observation in these simulations result in the additional time bins covering a much wider range of LSTs, rather than sampling a smaller range more finely. From \Cref{fig:duration_signal_results}, it can be seen that, again, the error ranges on the signals reduce as each additional day is included. However, it can also be seen that the error ranges are consistently smaller than those of the corresponding day separation, with a correspondingly larger DKL. As each pair of results uses data with the same number of time bins, this effect must arise from the improved foreground modelling made possible by the more complete observations of the foregrounds. This is again an expected result from the effects seen in \Cref{fig:cont_signal_results_hex} and \Cref{fig:cont_signal_results_ls}.

\begin{figure*}
\centering
\includegraphics[width=0.9\textwidth]{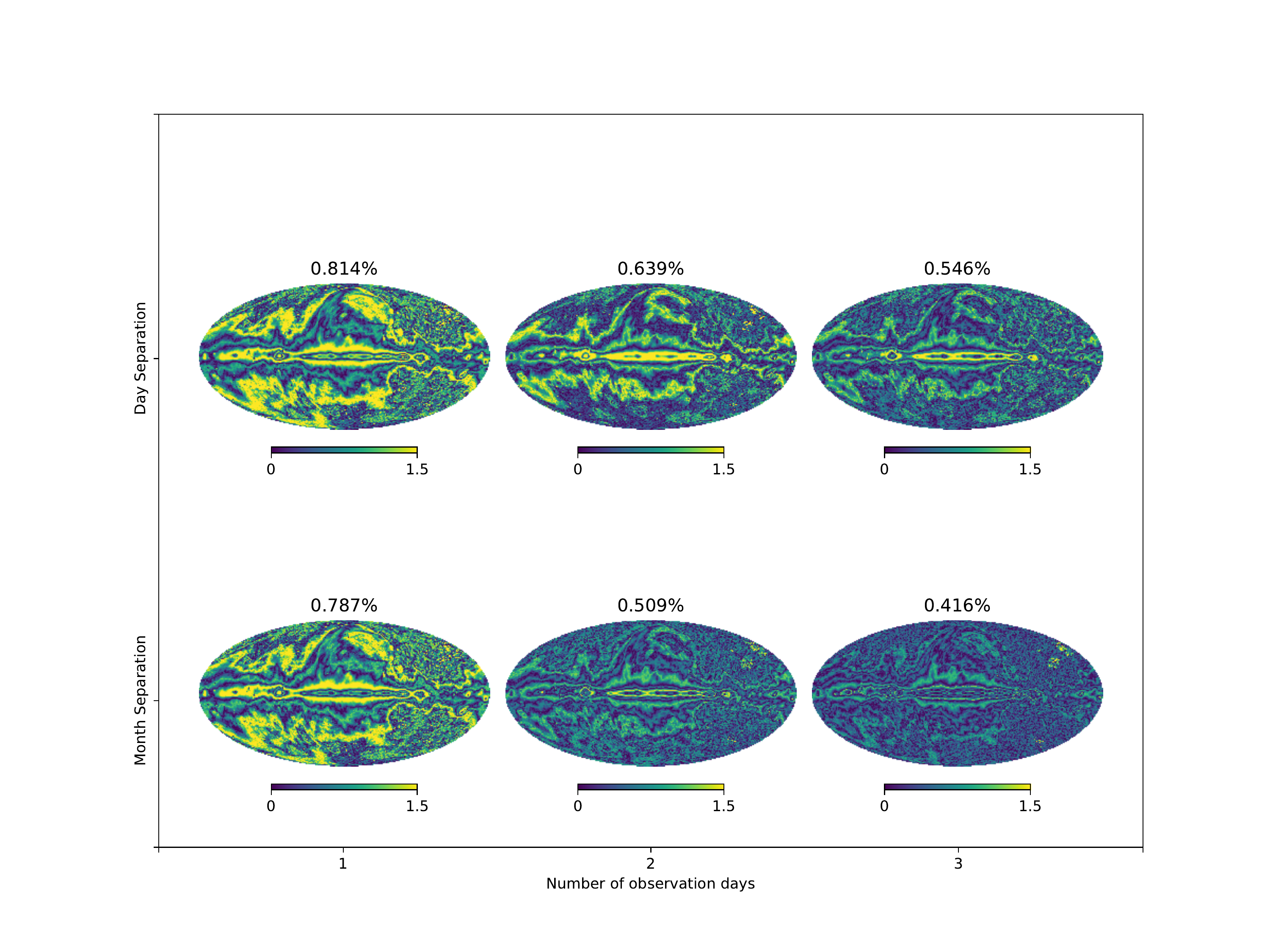}
\caption{Plots of the absolute percentage difference between the `true' spectral index map $\beta\left(\Omega\right)$ used to generate simulated data according to \Cref{eq:sim_data_gen} and the best fit spectral index maps for each simulated data set shown in \Cref{fig:duration_bins}, using the time separated likelihood \Cref{eq:sep_L}. In each case, the number of regions $N$ that the map is divided into is given by the number that gave the maximum Bayesian evidence, as shown in \Cref{fig:duration_peak_regions}, and the value of the spectral index in each region is given by the weighted posterior average of the fit parameters. The average percentage differences are recorded above each plot.}
\label{fig:duration_fore_maps}
\end{figure*}

In order to quantify this foreground recovery, the reconstructed foreground maps and their differences to the `true' map used to simulate the data were again calculated. These results are shown in \Cref{fig:duration_fore_maps}. As expected, both separations show an improvement in map accuracy as the number of observation days increases. The month separation case improves to a much greater degree and, in each case, gives a more accurate foreground reconstruction than the day separation, owing to the data sets sampling a much wider LST range and so containing a greater quantity of information about the foreground structure.

These results, therefore, confirm the expected behaviour, demonstrating that modelling foregrounds by fitting each data bin to a model on a joint set of parameters improves with longer observations. As additional nights of observation are included, both the recovered signal and the recovered foregrounds are seen to improve in accuracy.

\subsection{Fixed numbers of time bins}\label{variations_fixed_steps}
\begin{figure}
    \centering
    \includegraphics[width=\columnwidth]{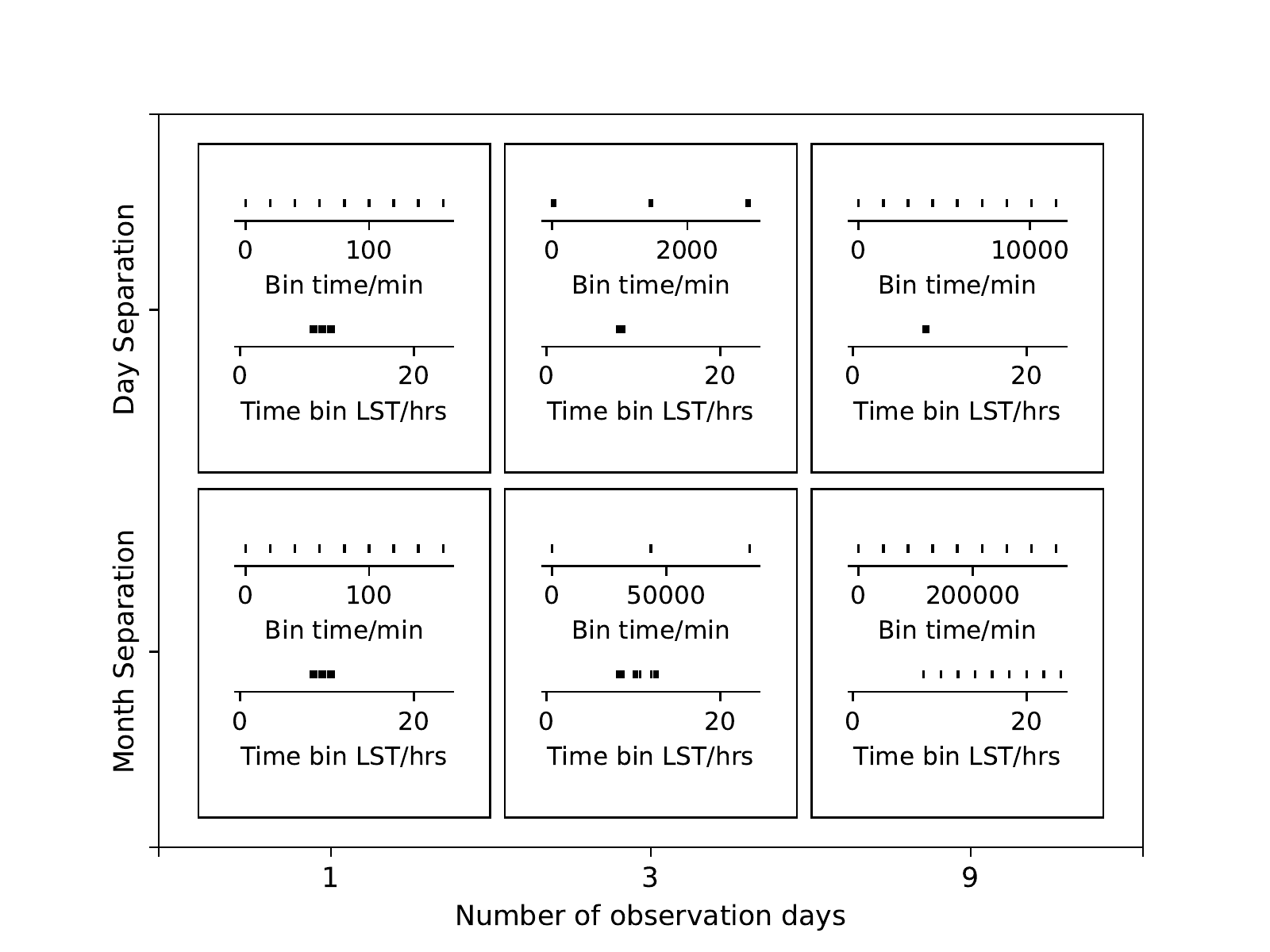}
    \caption{Time bin layouts for tests investigating the effects of observations with a constant number of time bins spanning different LST ranges. In each, 9 total time bins are used. The number of nights of the observations are spread over is shown on the x axis and the y axis showing the separation between the nights of observation, with `day separation' referring to 24 hours between night of observation and `month separation' referring to 720 hours (30 days) between each night of observation. On any given night, the observation begins at 00:00:00 and if there are multiple time bins on that night, they are separated by 20 minute intervals. In each subplot, the upper plot shows the time of each time bin after the start time of 00:00:00 01-01-2019 UTC and the lower plot shows the LSTs of those time bins.}
    \label{fig:steps_bins}
\end{figure}

It has been noted that the improvement in signal recovery seen when the number of time bins present in the data set increases is likely due to a combination of a reduction in the impact of noise and improved accuracy of foreground modelling due to the more complete foreground sampling. In order to investigate the relative impacts of these two effects, a new collection of simulated data sets was generated that all contained the same number of time bins, but spread over different LST ranges. Again, observations beginning at 00:00:00 01-01-2019 for an antenna at the Karoo radio reserve were assumed. The first 3 data sets consisted of 9 time bins, separated by 20 minute intervals, on a single night, 3 successive nights, each with 3 time bins separated by 20 minute intervals, beginning at 00:00:00 on each night, and a single bin at 00:00:00 on each of 9 successive nights. This results in a constant number of data bins, but covering a successively smaller LST range. As before, this will be referred to as `day separation'. The second 3 data sets followed the same pattern, but with 720 hours (30 days) separation between the cluster of bins, rather than 24 hours. This results in data sets of a constant number of time bins, but covering a progressively larger LST range. This will be referred to as `month separation'. The full layout of these time bins is shown in \Cref{fig:steps_bins}

\begin{figure}
    \centering
    \includegraphics[width=\columnwidth]{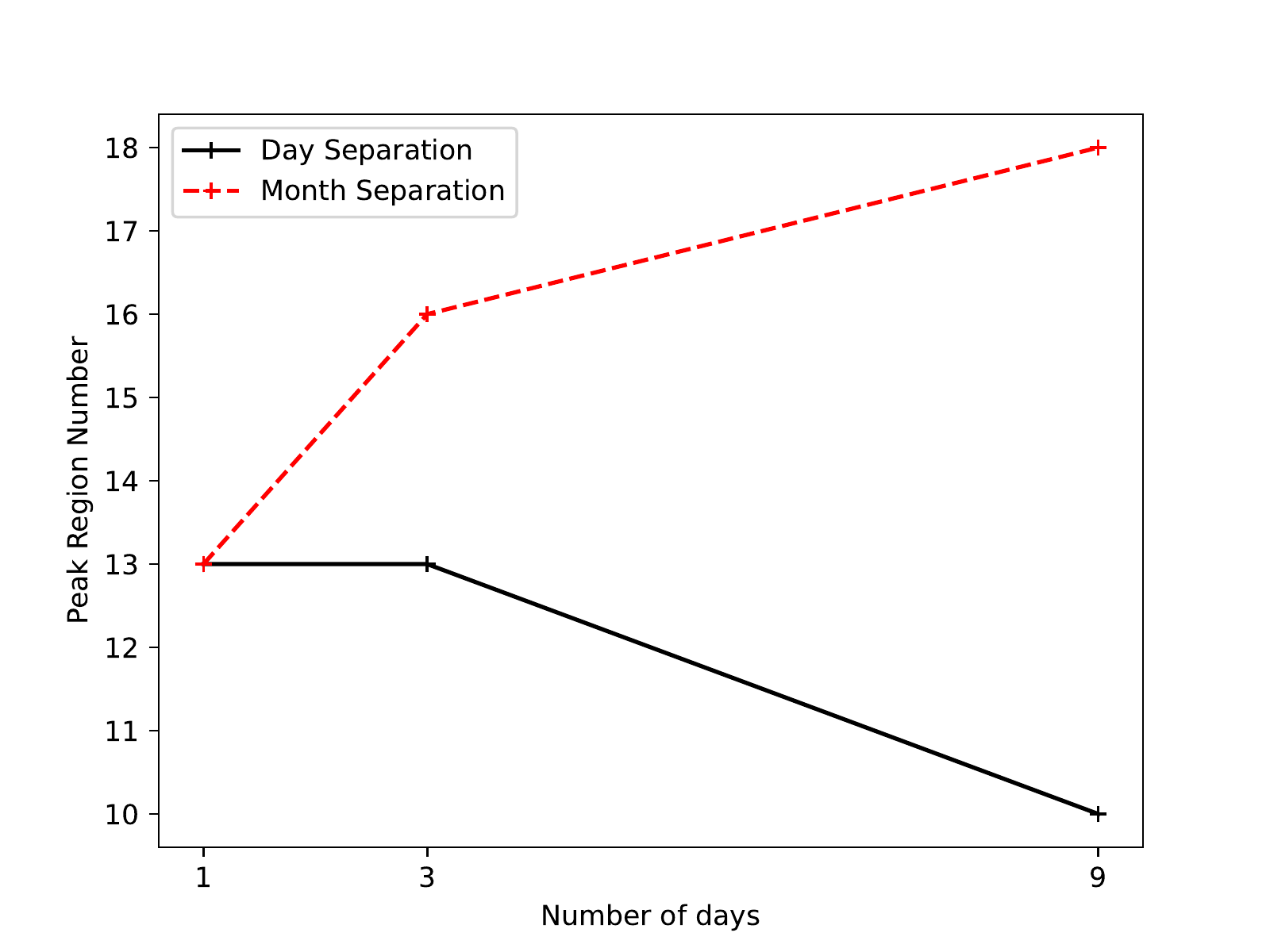}
    \caption{Plot of the number of regions $N$ that the foreground model was divided into in order to give the highest evidence model fit, using the time separated fitting method, for each of the simulated data sets shown in \Cref{fig:steps_bins}, generated using a log spiral antenna.}
    \label{fig:steps_peak_regions}
\end{figure}

As before, each of these data sets was fit using the time separated fitting procedure for a range of $N$s. \Cref{fig:steps_peak_regions} shows the values of $N$ that gave the optimum evidence in each case.

\begin{figure*}
\centering
\begin{subfigure}[b]{\columnwidth}
\includegraphics[width=\columnwidth]{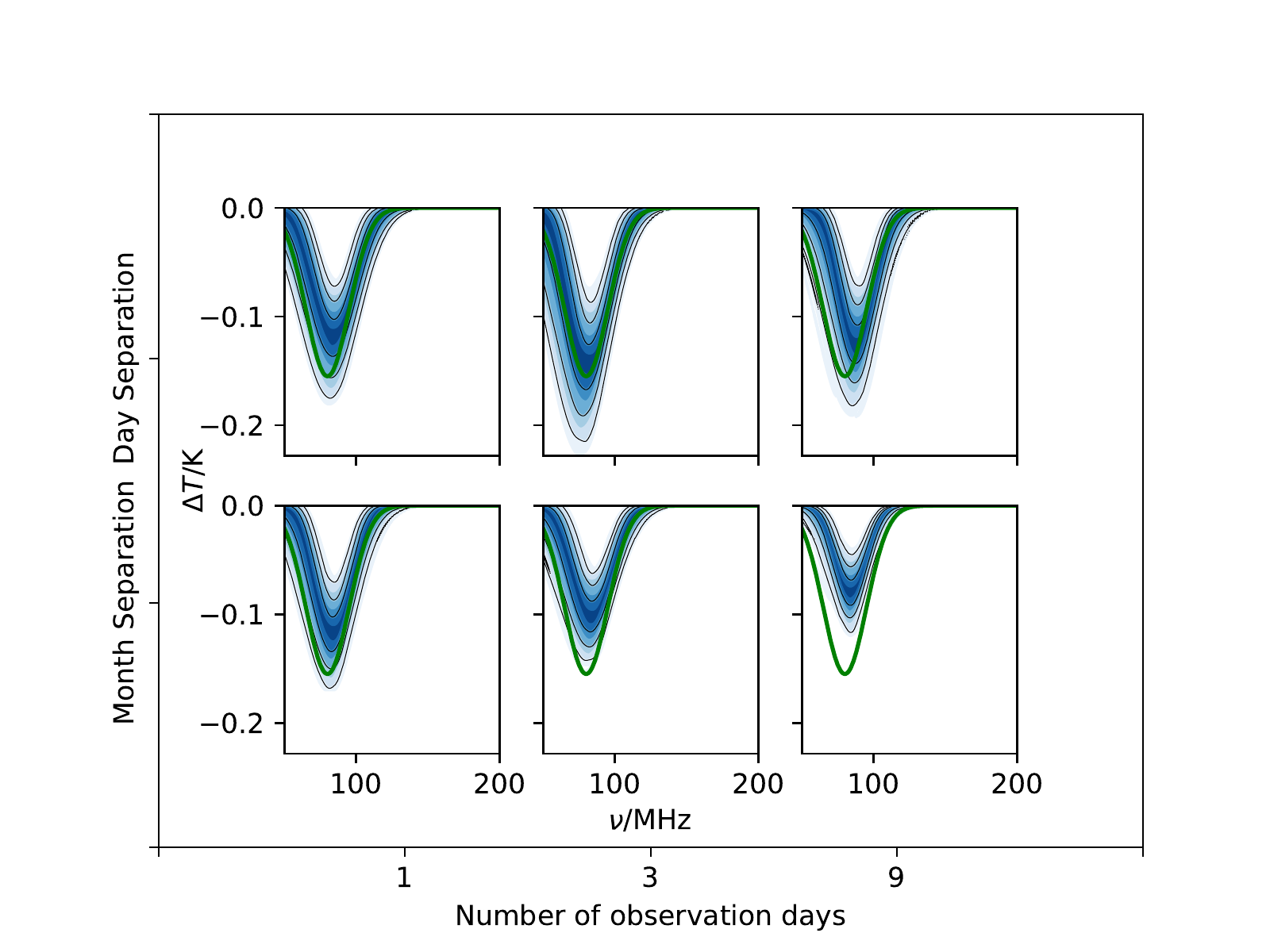}
\caption{Optimum evidence recovered signals}\label{fig:steps_op}
\end{subfigure}
\begin{subfigure}[b]{\columnwidth}
\includegraphics[width=\columnwidth]{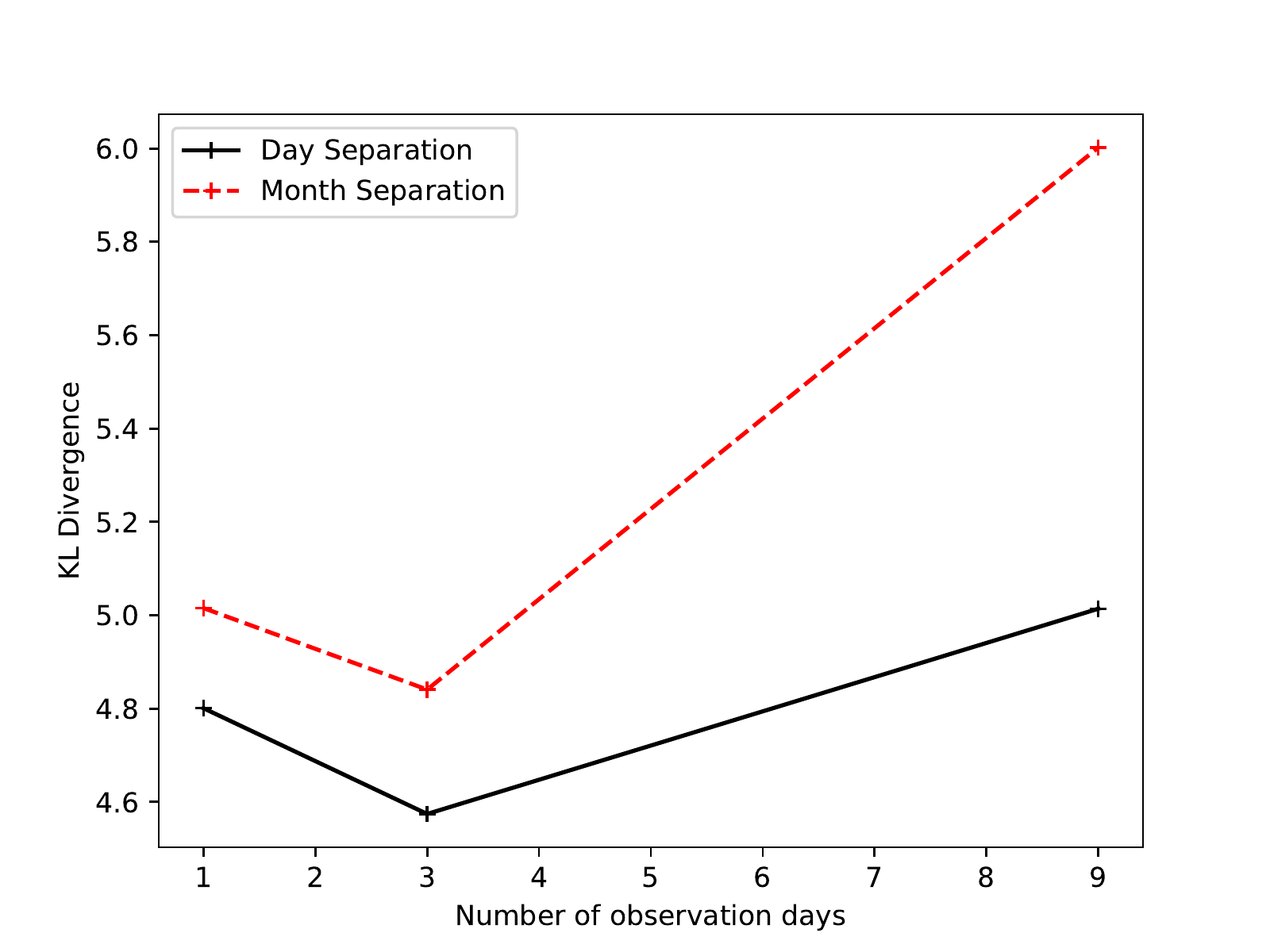}
\caption{Kullback–Leibler divergences of recovered signals}\label{fig:steps_op_KLD}
\end{subfigure}
\caption{Plots of the signal model fits to each simulated data set shown in \Cref{fig:steps_bins}, using the time separated likelihood \Cref{eq:sep_L}, for the number of foreground regions $N$ that gave the maximum Bayesian evidence, as shown in \Cref{fig:steps_peak_regions}. The left plots show the recovered signal model in each case in blue, with each colour band corresponding to a $\frac{1}{2}\sigma$ uncertainty on the signal model. The 'true' signal inserted into the simulated data is shown by a green line. The right plot shows the Kullback–Leibler divergences of the three signal parameters (centre frequency, width and amplitude) for the model fit to each data set.}
\label{fig:steps_signal_results}
\end{figure*}

\Cref{fig:steps_signal_results} shows the recovered signals and marginalised DKLs for each of these optimal $N$ fits.

Considering first the day separation case, it can be seen that there are only slight changes to the signal recovery and the DKLs with changing LST spread of the data bins. This implies that the reduction of the impact of noise is the more dominant effect in these cases. However, if the month separation cases are also considered, they show consistently higher DKLs and smaller error ranges on the signal recovery. As these month separated data sets contain the same number of data bins, but increase in LST range rather than decrease, this demonstrates that changing the LST range covered for a fixed number of data bins has an increasing impact on the signal recovery as the LST range covered becomes larger, but the effect is small compared to that of reduction in noise impact when the LST range is small.

\begin{figure*}
\centering
\includegraphics[width=0.9\textwidth]{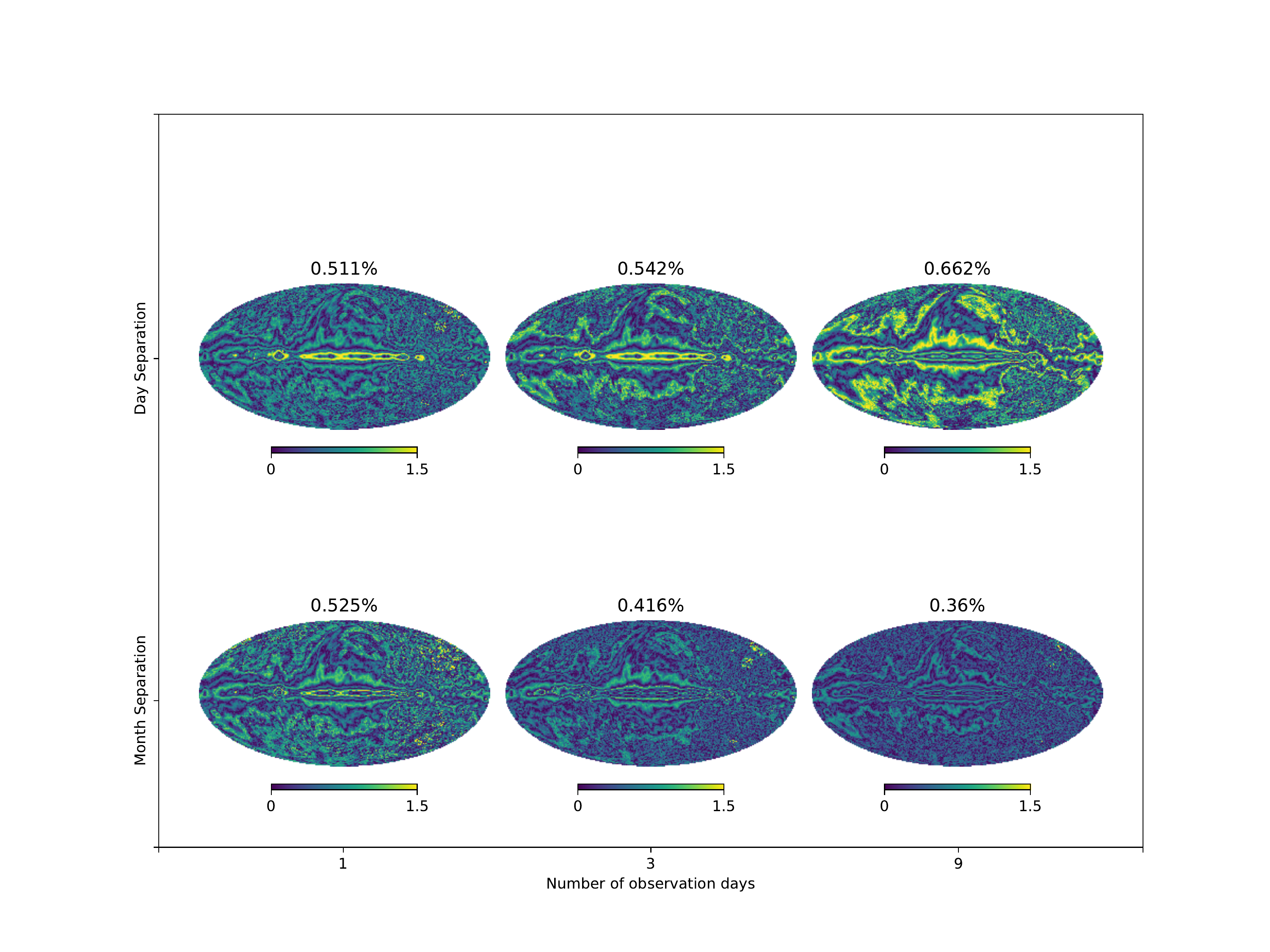}
\caption{Plots of the absolute percentage difference between the `true' spectral index map $\beta\left(\Omega\right)$ used to generate simulated data according to \Cref{eq:sim_data_gen} and the best fit spectral index maps for each simulated data set shown in \Cref{fig:steps_bins}, using the time separated likelihood \Cref{eq:sep_L}. In each case, the number of regions $N$ that the map is divided into is given by the number that gave the maximum Bayesian evidence, as shown in \Cref{fig:steps_peak_regions}, and the value of the spectral index in each region is given by the weighted posterior average of the fit parameters. The average percentage differences are recorded above each plot.}
\label{fig:steps_fore_maps}
\end{figure*}

\Cref{fig:steps_fore_maps} shows the errors in the foreground reconstructions of these model fits. It can be seen that, as before, the wider the range of LSTs samples by the data, the more accurate the resulting reconstructed foreground map. This effect is even seen in the day separation case, in which the change in quality of the signal recovery was small. This demonstrates that, while the signal recovery does not benefit a great deal from small changes in LST range covered without additional data bins when performing a simultaneous model fit, the foreground reconstruction still benefits significantly. 

Another feature that can be noted in the results of \Cref{fig:steps_signal_results} is that the case of the 9 time bins each being on an individual night 30 days apart shows a high DKL, but a recovered signal that significantly underestimates the amplitude of the true signal. This effect appears to be due to differences in the parameters being fit for by the different time bins. The process of time separated model fitting being discussed here is dependent on each model being fit to each data bin converging on the same parameter values. Typically, this is the case for the A21 foreground model.  However, the model is, by definition, only an approximation to the true spectral index distribution of the sky. The fit spectral index parameters, therefore, do not have theoretical `true' values, although they converge towards this as $N$ increases. As a result, if two time bins each observe only part of the sky for a given region, they may converge on a slightly different value of the spectral index in that region. 

\begin{figure*}
\centering
\begin{minipage}{0.35\textwidth}
    \begin{minipage}{\textwidth}
         \begin{subfigure}[b]{\linewidth}
        \includegraphics[width=\textwidth]{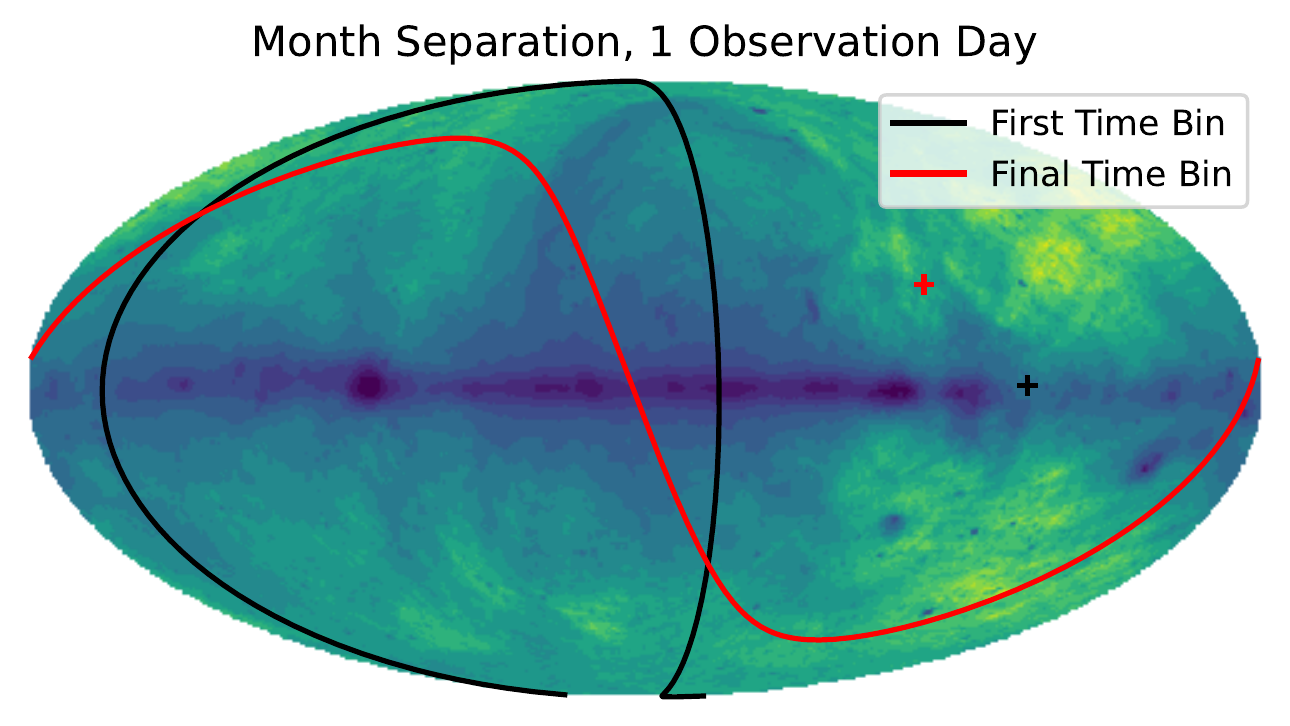}
        \end{subfigure}       
    \end{minipage}
  \par \smallskip
    \begin{minipage}{\textwidth}
        \begin{subfigure}[b]{\linewidth}
        \includegraphics[width=\textwidth]{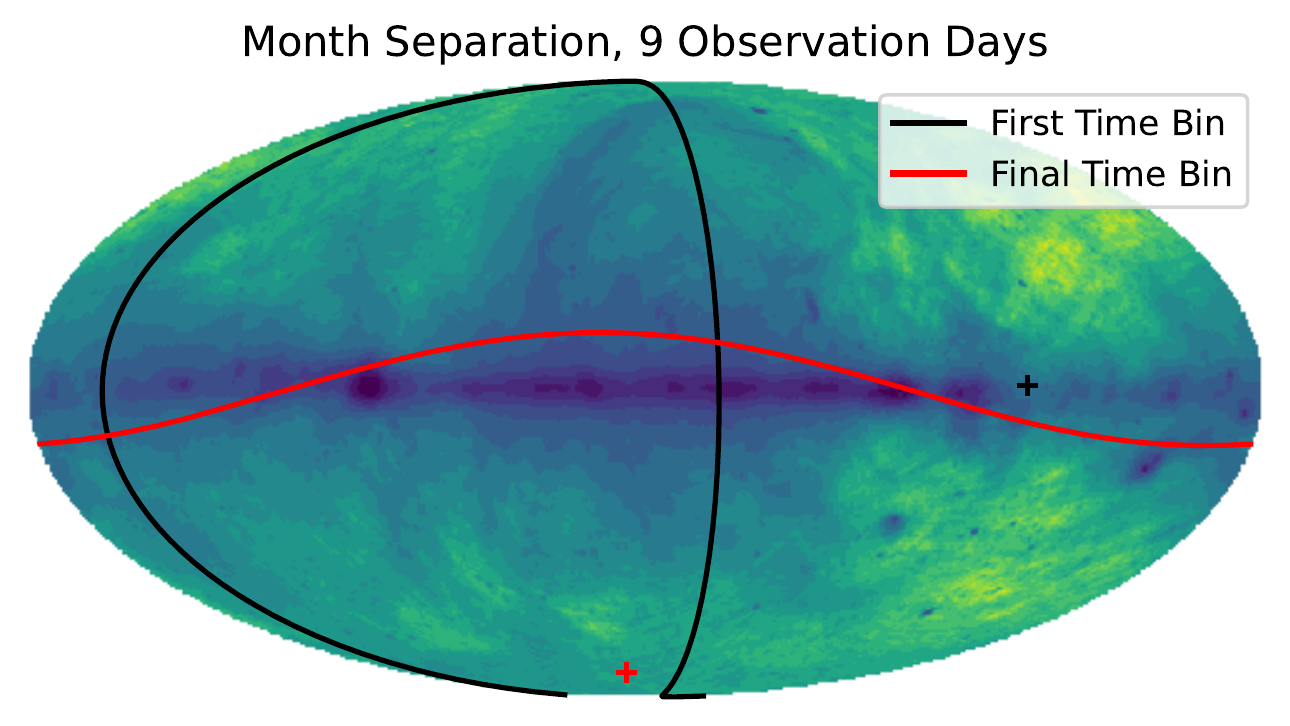}
        \end{subfigure}
    \end{minipage}
\end{minipage}
\begin{minipage}{0.55\textwidth}
    \begin{subfigure}[b]{\textwidth}
    \includegraphics[width=\linewidth]{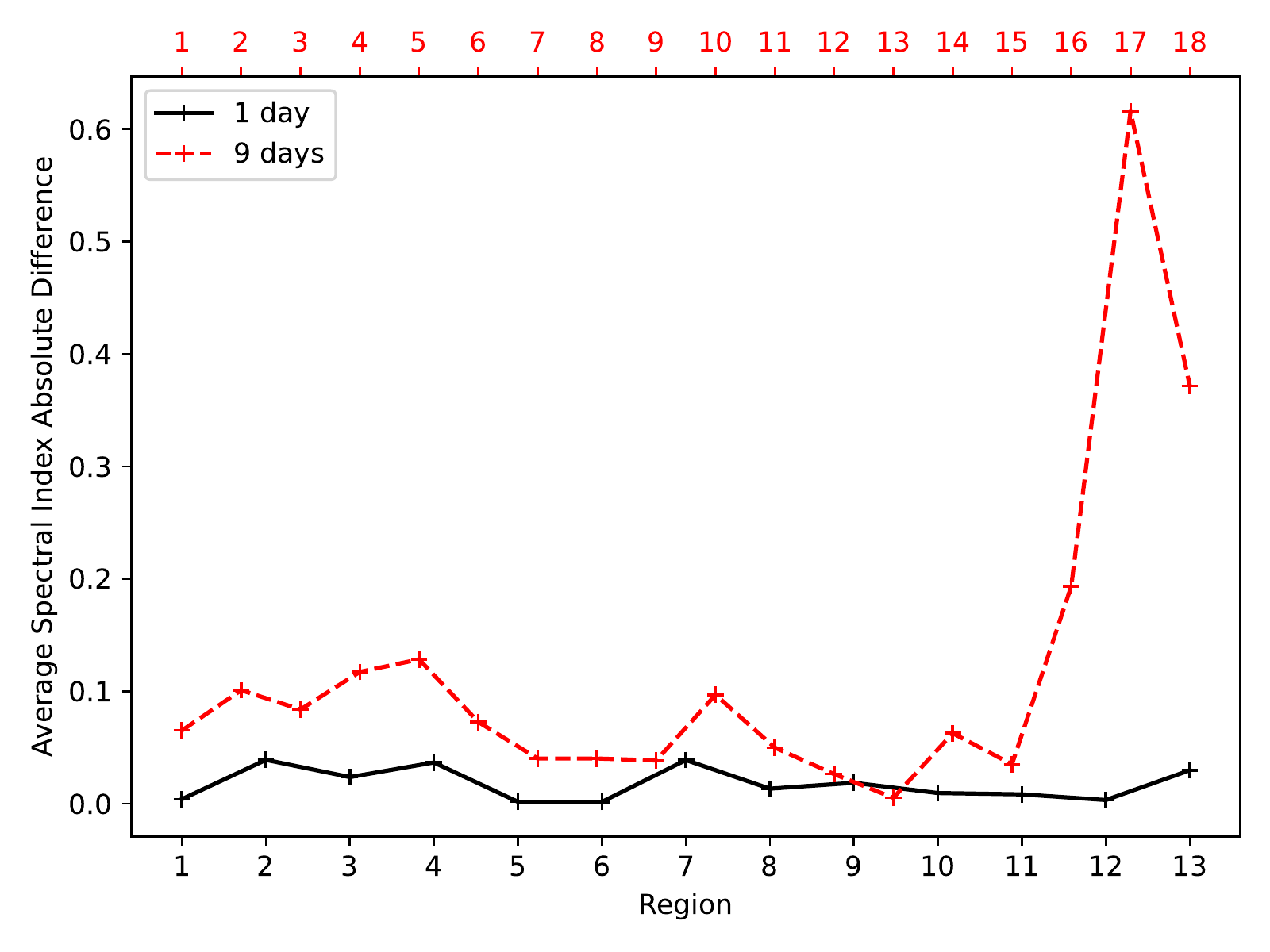}
    \end{subfigure}
\end{minipage}
\caption{Plots of the differences in spectral index between the sections of each region observed by an antenna at different times. The upper plots show, for the month separation, with 1 observing day case on the left and 9 observing days on the right, the parts of each region visible to the antenna for different time bins. In each case, the subdivision of the sky into the optimal number of regions, 13 for 1 day and 18 for 9 days, as specified in \Cref{fig:steps_peak_regions}, is shown, with the solid lines showing the horizon and the `+' showing the zenith for an antenna in the Karoo radio reserve at the first and last observing times, in black and red respectively. The lower plot shows the absolute differences between the average spectral indices of the parts of each region observed in the first and last time bins, as specified in the upper plots, weighted by the log spiral antenna beam gain at 50MHz. These differences are shown for the 1 day case in black and the 9 day case in red.}
\label{fig:param_diff}
\end{figure*}

This effect is demonstrated in \Cref{fig:param_diff}. The upper two plots show, for the 1 day and 9 day month separation cases, defined in \Cref{fig:steps_bins}, respectively, the parts of each sky region being observed in the first and last time bins. In each case, the peak number of regions shown in \Cref{fig:steps_peak_regions} is used. It can be seen that in each case, the different time bins observe different parts of some of the regions, with this difference being much larger for the 9 day case. As the `true' spectral index used to generate the data is not strictly uniform within each region, this may result in slightly different parameter values being favoured for different time bins. This difference can be quantified by calculating the average spectral index, weighted by beam gain, for the \emph{part} of each region being observed at each time. The lower plot of \Cref{fig:param_diff} shows the absolute differences between these averages for the first and last time bins of the month separation 1 day and 9 day cases. In each case, the spectral indices were weighted by the log spiral beam gain at 50MHz. It can be seen that the differences in the region spectral indices between time bins, and so the parameter values the pipeline is trying to fit for, are consistently much larger for the 9 day case. As was seen in earlier results, if the sky is thoroughly sampled, this effect averages out. However, in a case such as this, with a relatively small number of data bins spread over a very wide LST range, this may result in differences between the parameters of each time bin model, which in turn distorts the signal somewhat. This demonstrates that it is important when using this process for the data to be time-binned finely enough to thoroughly sample the foregrounds.

\section{Multiple Antennae}\label{sec:multi_ant}
The concept of simultaneous fitting of multiple data sets at different observing times to corresponding models in a single joint likelihood discussed here is dependent on using a model in which a physical property is used for the parameters. Such a model ensures that data at different observing times is fit by the same parameter values. We have demonstrated that this property can be used to gain additional information about the structure of the foregrounds from their time rotation, allowing more accurate modelling of both the foregrounds and the 21cm signal than could be achieved by time-averaging data. 

This concept can, however, be extended further. It can be noted that in the modelling process proposed in A21, the parameters of the foreground model are not only independent from time rotation, but also from the antenna used in the observation. This raises the possibility that, as well as fitting data sets from multiple observing times jointly in the same likelihood to the same parameters, data from multiple different antennae could also be fit jointly in this manner. Analogously to how joint fitting of data sets from different observing times exploits the time rotation of the foregrounds to achieve more accurate modelling, joint fitting of data from multiple antennae would potentially exploit the different chromatic distortion patterns of the different antennae in order to improve modelling of the foregrounds.

In order to perform such a joint fit in practice in a Bayesian model fitting, the likelihood in \Cref{eq:sep_L} should be expanded to 

\begin{multline}\label{eq:sep_L_ant}
    \log\mathcal{L} = \sum_{i}\sum_{j}\sum_{k}-\frac{1}{2}\log\left(2\pi\sigma_\mathrm{n}^{2}\right) \\ - \frac{1}{2}\left(\frac{{T_\mathrm{data}}_k\left(\nu_{i}, t_{j}\right)-\left({T_\mathrm{F}}_k\left(\nu_{i}, t_{j}, \theta_\mathrm{F}\right) + T_{S}\left(\nu_{i},\theta_\mathrm{S}\right)\right)}{\sigma_\mathrm{n}}\right)^{2},
\end{multline}
where ${T_\mathrm{data}}_k\left(\nu_{i}, t_{j}\right)$ refers to observation data at frequency $nu_i$ and time $t_j$ made using antenna $k$ and ${T_\mathrm{F}}_k\left(\nu_{i}, t_{j}, \theta_\mathrm{F}\right)$ refers to a corresponding foreground model for the same frequency, time and antenna.

\begin{figure*}
\centering
\begin{subfigure}[b]{\columnwidth}
\includegraphics[width=\columnwidth]{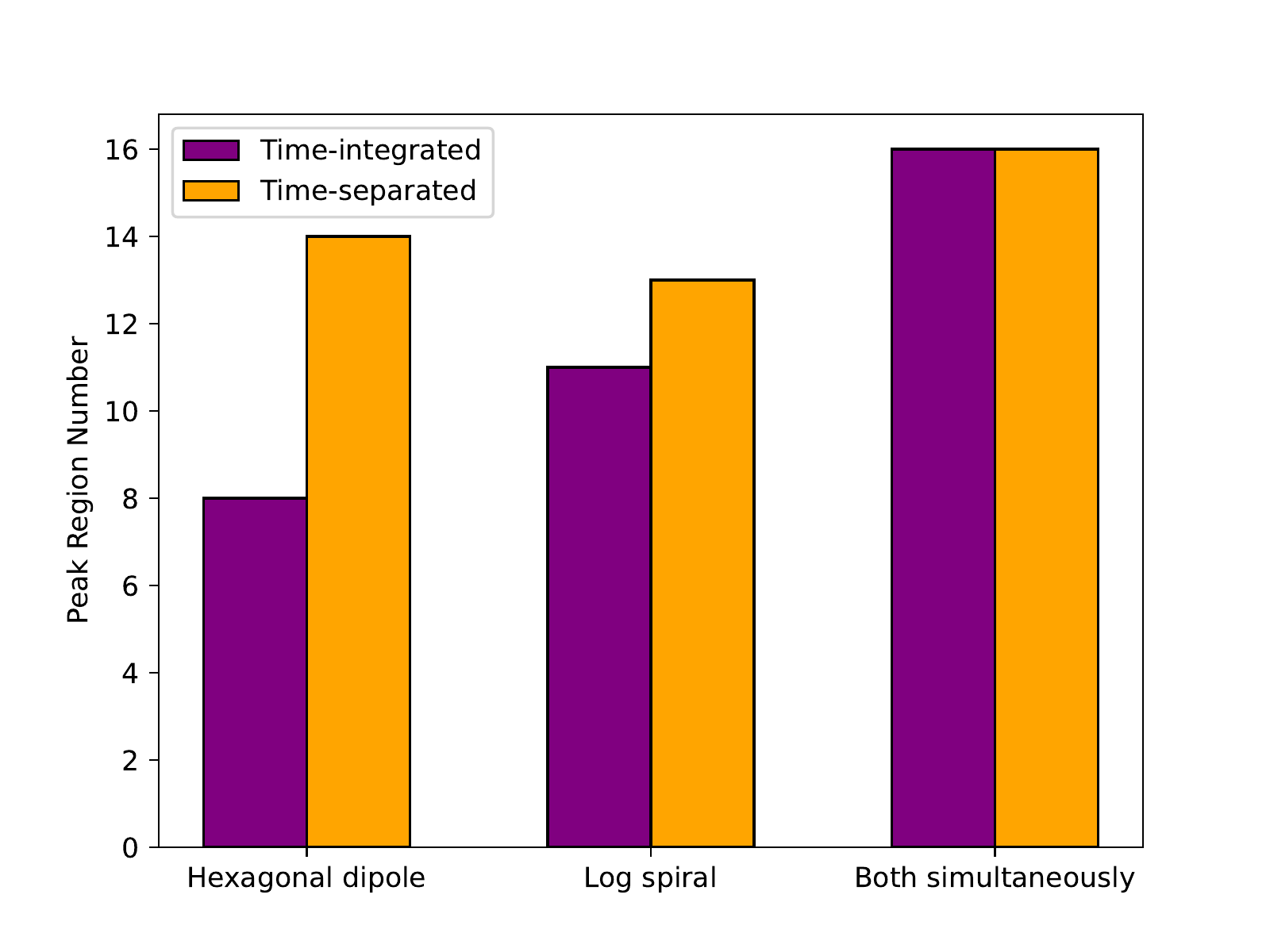}
\caption{$N$ that gives optimal Bayesian evidence}\label{fig:ant_peaks}
\end{subfigure}
\begin{subfigure}[b]{\columnwidth}
\includegraphics[width=\columnwidth]{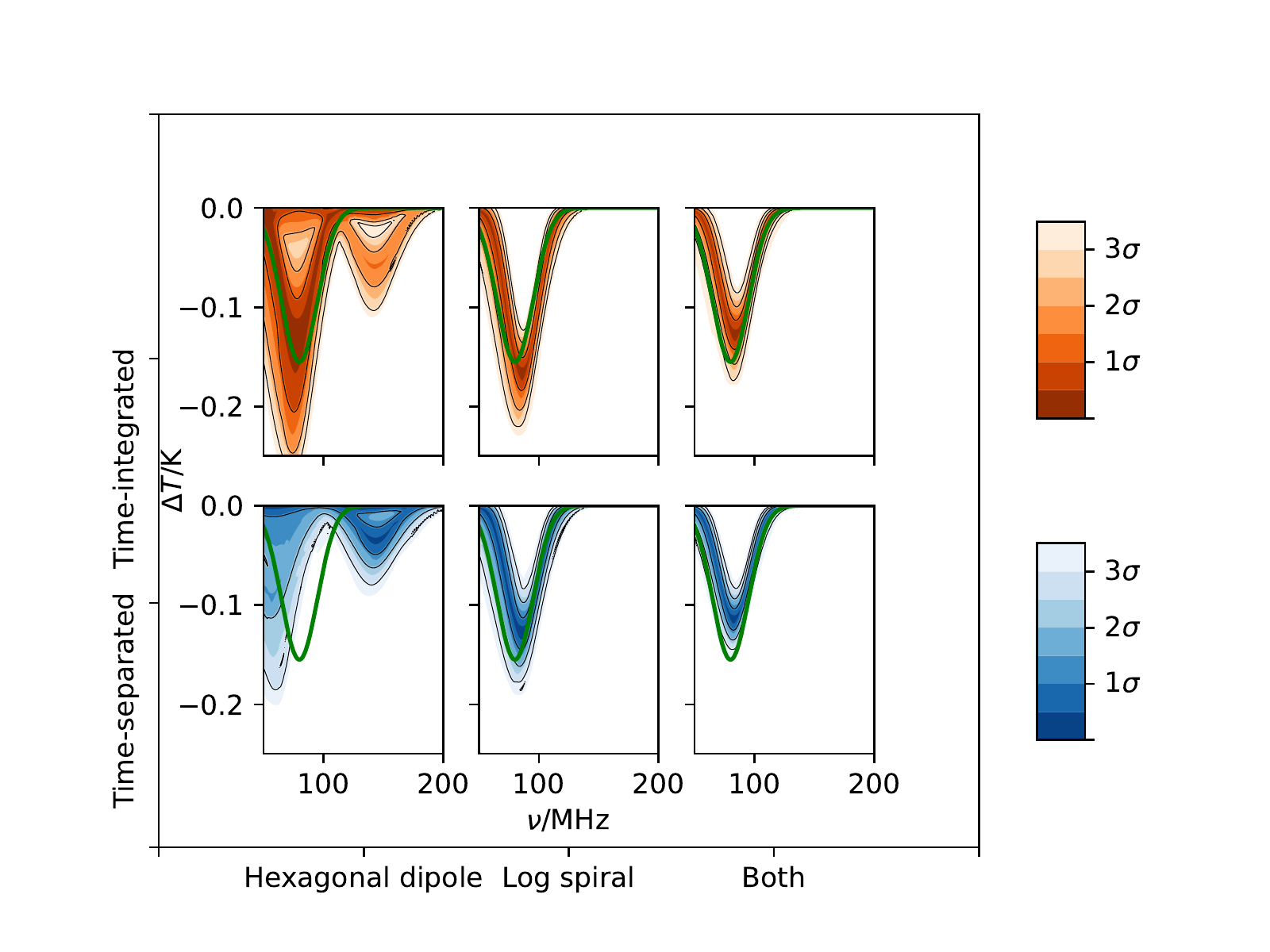}
\caption{Optimum evidence recovered signals}\label{fig:ant_op}
\end{subfigure}
\begin{subfigure}[b]{\columnwidth}
\includegraphics[width=\columnwidth]{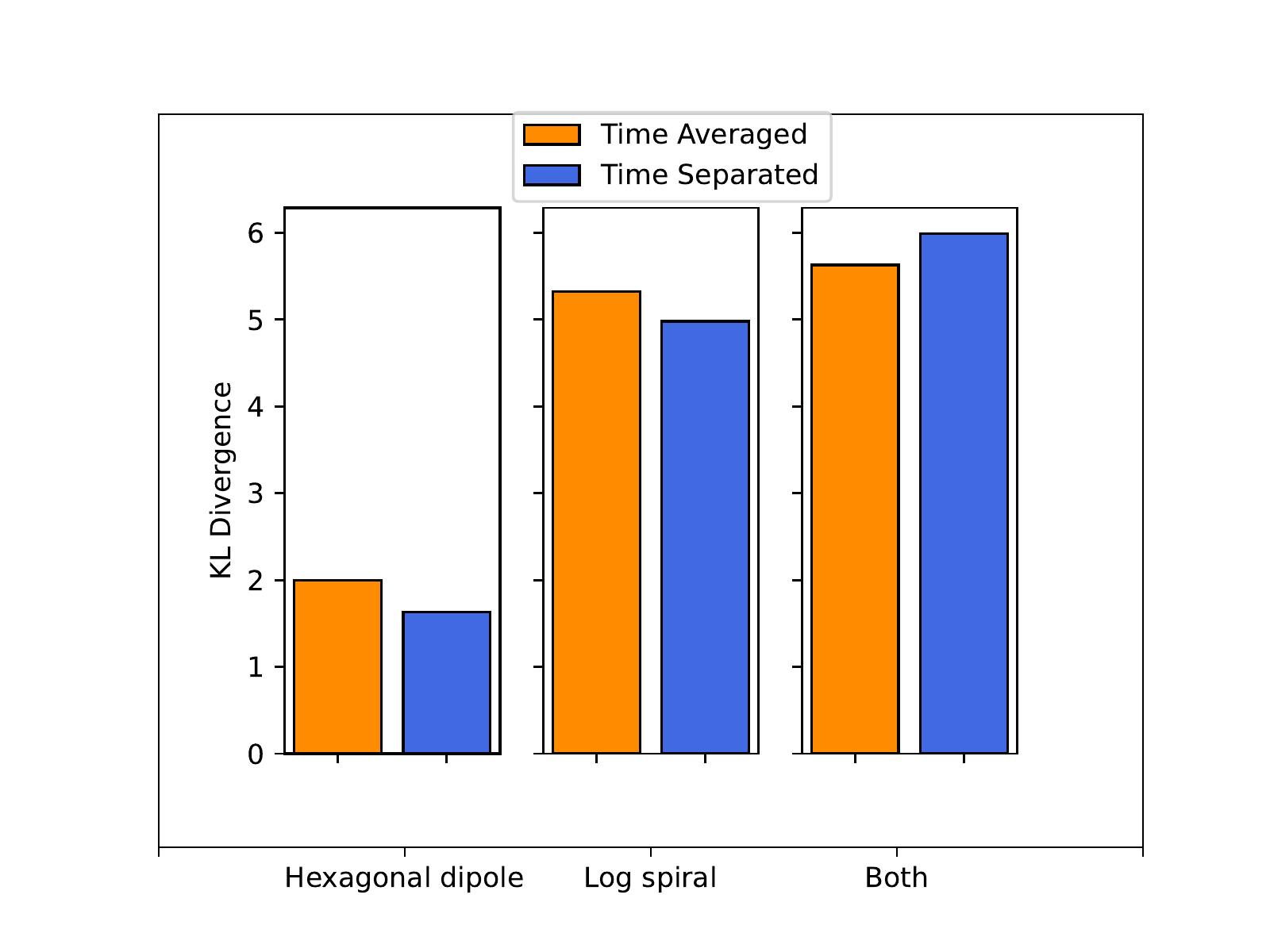}
\caption{Kullback–Leibler divergences of recovered signals}\label{fig:ant_op_KLD}
\end{subfigure}
\caption{Plots of the signal model fits to simulated data sets with 1 hour of observation and 5 minutes between data bins, made with a hexagonal dipole antenna and a log spiral antenna. Each plot shows the results of fitting the data sets of the two antenna separately in a time integrated sense according to \Cref{eq:integ_L_exp} and a time-separated sense according to \Cref{eq:sep_L}, as well as jointly, according to \Cref{eq:sep_L_ant}. \Cref{fig:ant_peaks} shows the number of regions required to give the maximum evidence, \Cref{fig:ant_op} shows the recovered signal models for these optimal foreground models, compared to the `true' signal inserted into the data shown in green, and \Cref{fig:ant_op_KLD}, shows the marginalised DKLs of these signal models.}
\label{fig:multi_antenna}
\end{figure*}

\Cref{fig:multi_antenna} shows the results of fitting simulated data sets from a log spiral and a hexagonal dipole antenna jointly in this manner, in comparison to when the data sets from each antenna are fit alone. For this example, we used the data set of a single hour of integration and 5 minute intervals between time bins. This is an intermediate case, with the log spiral antenna able to recover the signal, but not the hexagonal dipole, which was chosen to investigate if data in which a signal could not be detected in directly could produce any benefit if fit jointly with data from another antenna.

From \Cref{fig:multi_antenna}, it can be seen that simultaneous fitting of data from two antennae does produce a reduction in uncertainty on the reconstructed signal, both for time-integrated and time-separated data. A corresponding increase in DKL is also seen, particularly in the time separated case.

\begin{figure*}
\centering
\includegraphics[width=0.9\textwidth]{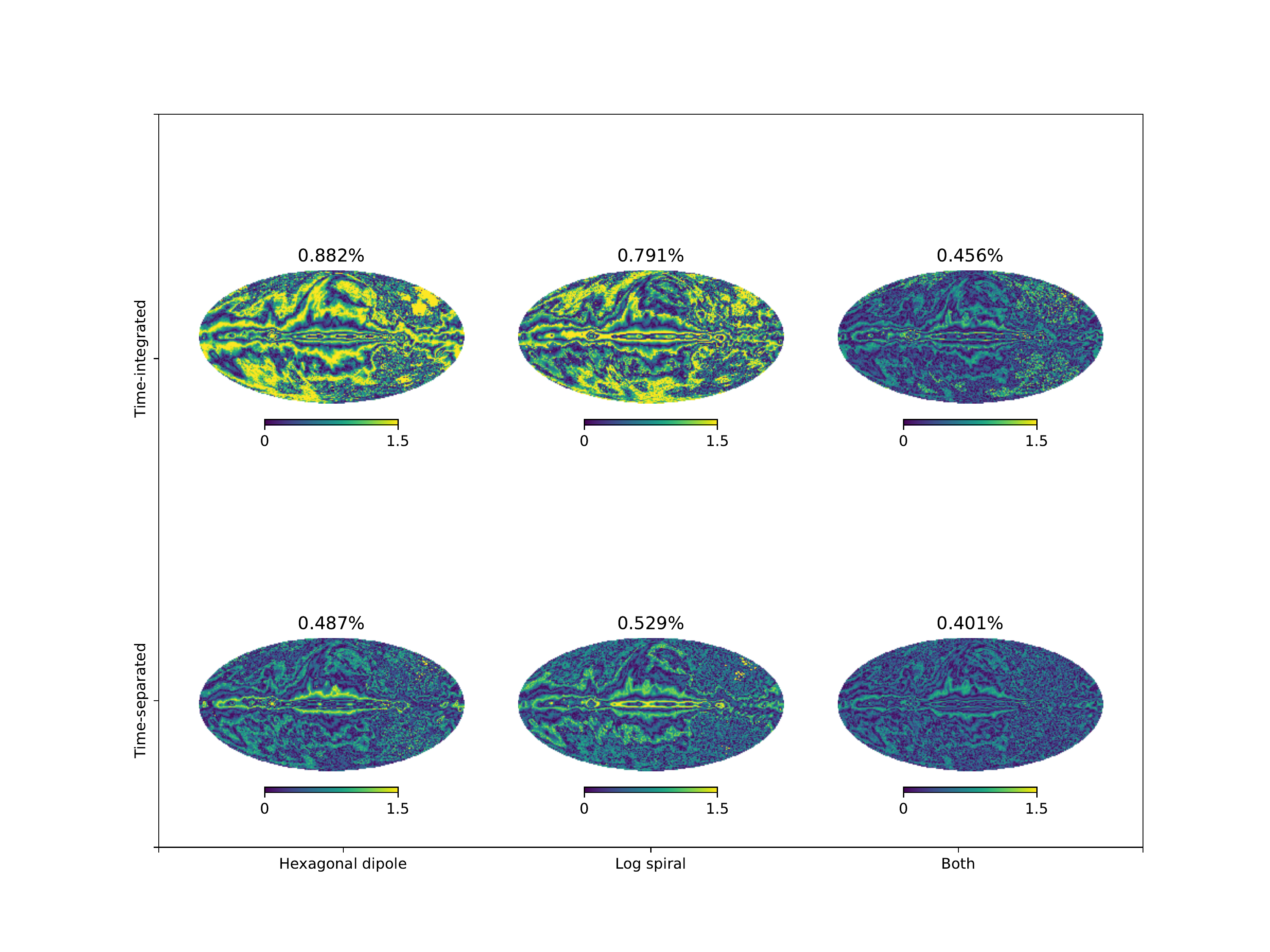}
\caption{Plots of the absolute percentage difference between the `true' spectral index map $\beta\left(\Omega\right)$ used to generate simulated data according to \Cref{eq:sim_data_gen} and the best fit spectral index maps for simulated data sets of the two antenna separately, fit in a time integrated sense according to \Cref{eq:integ_L_exp} and a time-separated sense according to \Cref{eq:sep_L}, as well as jointly, according to \Cref{eq:sep_L_ant}. In each case, the number of regions $N$ that the map is divided into is given by the number that gave the maximum Bayesian evidence, as shown in \Cref{fig:ant_peaks}, and the value of the spectral index in each region is given by the weighted posterior average of the fit parameters. The average percentage differences are recorded above each plot.}
\label{fig:multi_antenna_foreground}
\end{figure*}

\Cref{fig:multi_antenna_foreground} shows the errors in the fitted foreground spectral index maps compared to the `true' map used to generate the simulated data. It can be seen that fitting the data with both antenna simultaneously results in a much more accurate map than can be achieved with either antenna data set alone. This shows, therefore, that by using data from multiple antennae, the extra information about the nature of the foregrounds and chromatic distortions that is gained from how the distortions change between antennae allows the foregrounds to be modelled much more accurately and, as a result, also allows a more precise signal reconstruction.

It can also be seen from \Cref{fig:multi_antenna_foreground} that the difference in foreground accuracy between the time-averaged and time-separated cases for the joint fit is much smaller than that of each individual antenna. This implies that the extra information about the foregrounds gained from taking data with multiple different antennae is greater than that gained from changes in the foregrounds with time.

\section{Conclusions}\label{sec:conclusions}
The ability to accurately model bright radio foregrounds and chromatic distortions from antennae is highly important for achieving an accurate detection of the global 21cm signal by a global experiment. In this paper, we explored how the parameterisation used in the A21 pipeline is based on physical properties that should be unchanged with the rotation of the foregrounds with time enables data sets from multiple times to be fit simultaneously with a single Bayesian likelihood.

We demonstrated that for a dipole antenna like that being used in REACH \citep{mission}, this simultaneous fitting process enables the global 21cm signal to be recovered more accurately and precisely than if the time-average of these data sets is fitted, owing to the additional information about the foregrounds and chromaticity obtained from knowledge of their changes with the Earth's rotation. Furthermore, we showed that this simultaneous fitting process enables the foregrounds to be reconstructed to a much higher degree of accuracy, demonstrating this information gain about the foregrounds.

We also showed that for a less chromatic, easier to model, log spiral antenna, the improvement in signal recovery was a lot less significant, owing to the signal recovery being a lot less limited by lack of foreground and chromaticity information for this antenna. However, a significant improvement is foreground modelling was still seen, showing this technique still has benefits for foreground modelling.

We tested the effects of changing the number of time bins used and their spread in LST and found that, in general, the more data sets being fit simultaneously the better both the signal and foreground recovery are, partially due to reduced effect of noise and partially due to additional foreground information.
This is in agreement with the findings of \citet{tauscher20b}, which showed that including multiple time bins of data simultaneously gave a reduction in errors for the SVD foreground modelling pipeline of \citet{tauscher18}.

Furthermore, signal and foreground recovery were found to improve in general with increasing LST range covered by the data sets. We also identified a limitation with this technique, in which having too few data bins separated by too large time gaps can result in each time bin requiring different parameters, which can bias the recovered signal.

Finally, we demonstrated how this simultaneous modelling technique can be extended to simultaneous fitting of data from multiple different antennae, taking advantage of the different chromaticity structures to identify and model them more accurately. We showed that by fitting data from both a hexagonal dipole and log spiral antenna simultaneously, lower errors are achieved in both the foreground and signal recovery to a much greater degree than from using time separation.

The time-separated data set fitting procedure discussed here, as well as improving both signal and foreground modelling directly, could potentially also be beneficial for other aspects of 21cm experiment data analysis. In particular, fitting multiple data sets from different times to the same parameter values may help to avoid the effect of time dependent distortions in 21cm data, such as might arise from the ionosphere or uncorrected for RFI. These effects will be explored further in future work, with Shen et al. (in prep.) discussing the effects on ionospheric modelling.

\section*{Acknowledgements}
We would like to thank Quentin Gueuning and John Cumner for providing the electromagnetic simulations of the log spiral antenna and hexagonal dipole antenna respectively. We would also like to thank Harry Bevins for providing the analysis software MARGARINE, used to calculate DKLs.
Dominic Anstey and Eloy de Lera Acedo were supported by the Science and Technologies Facilities Council. Will Handley was supported by a Royal Society University Research Fellowship. We would also like to thank the Kavli Foundation for their support of REACH.

\section*{Data Availability}
The data underlying this article is available on Zenodo, at 10.5281/zenodo.7153212.




\bibliographystyle{mnras}
\bibliography{bibliography}





\bsp	
\label{lastpage}
\end{document}